\documentclass{article}

\usepackage[preprint]{neurips_2025}

\usepackage[utf8]{inputenc} 
\usepackage[T1]{fontenc}    
\usepackage{hyperref}       
\usepackage{url}            
\usepackage{booktabs}       
\usepackage{amsfonts}       
\usepackage{nicefrac}       
\usepackage{microtype}      
\usepackage{xcolor}         
\usepackage{amsmath}
\usepackage{booktabs} 
\usepackage{multirow} 
\usepackage{array}    

\usepackage{tcolorbox}
\usepackage{listings}

\usepackage{makecell}
\usepackage{graphicx}
\usepackage{multirow}

\usepackage{color}

\usepackage{hyperref}
\usepackage{xcolor}

\hypersetup{
    colorlinks=true,      
    urlcolor=cyan,         
    pdfborder={0 0 0}  
}

\title{HeartMuLa: A Family of Open Sourced Music Foundation Models}

\author{%
   \textmd{HeartMuLa Teams}\\
  \small
    GitHub: \href{https://github.com/HeartMuLa/heartlib}{https://github.com/HeartMuLa/heartlib}\\
    Demo: \href{https://heartmula.github.io/}{https://heartmula.github.io/}
}

\begin{document}

\maketitle

\vspace{-2em}

\begin{abstract}
We present a family of open-source Music Foundation Models designed to advance large-scale music understanding and generation across diverse tasks and modalities. Our framework consists of four major components: (1) HeartCLAP, an audio-text alignment model; (2) HeartTranscriptor, a robust lyric recognition model optimized for real-world music scenarios; and (3) HeartCodec, a low-frame-rate (12.5 Hz) yet high-fidelity music codec tokenizer that captures long-range musical structure while preserving fine-grained acoustic details and enabling efficient autoregressive modeling; (4) HeartMuLa, an LLM-based song generation model capable of synthesizing high-fidelity music under rich, user-controllable conditions (e.g., textual style descriptions, lyrics, and reference audio). In addition, it provides two specialized modes: (i) fine-grained musical attribute control, which allows users to specify the style of different song sections (e.g., intro, verse, chorus) using natural language prompts; and (ii) short, engaging music generation, which is suitable as background music for short videos. Lastly, HeartMuLa improves significantly when scaled to 7B parameters. For the first time, we show that a Suno-level, commercial-grade system can be reproduced using academic-scale data and GPU resources.
We expect these foundation models to serve as strong baselines for future research and to facilitate practical applications in multimodal content production.

\end{abstract}

\begin{figure}[htbp]
    \centering
    \includegraphics[width=0.9\linewidth]{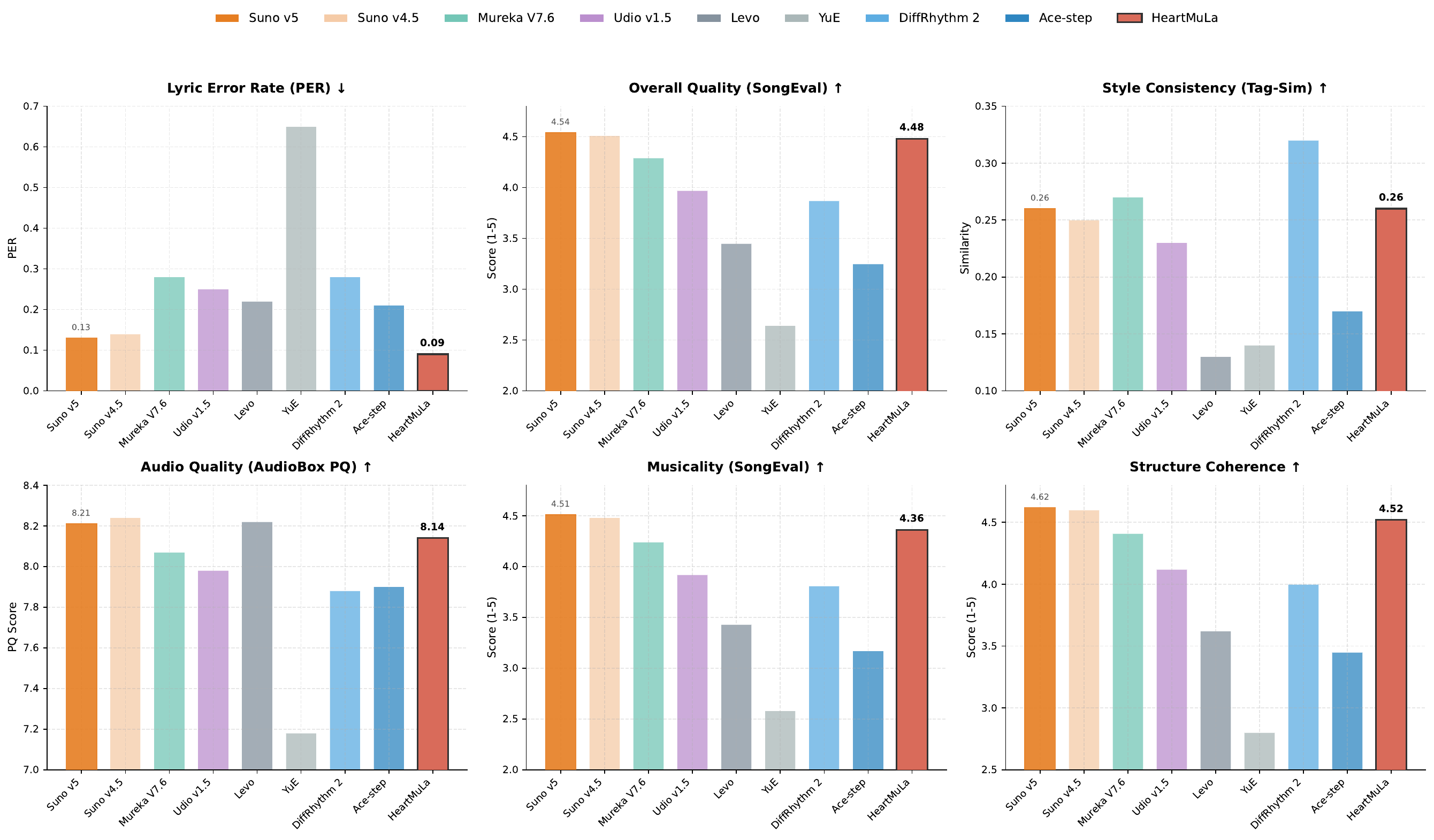}
    \caption{Overall comparison of HeartMuLa-oss-3B with existing music foundation models.}
    \label{fig:comparison}
\end{figure}

\newpage
\maketitle        
\tableofcontents  
\newpage

\section{Introduction}
Music generation and understanding have rapidly evolved with the emergence of large-scale multimodal foundation models~\cite{musicgen,levo,wu2025clamp3}. Recent advances in audio representation learning~\cite{mucodec_muencoder,musicfm}, text-audio alignment~\cite{elizalde2023clap}, and autoregressive music generation~\cite{yang2023uniaudio,musicgen,musiclm} as well as diffusion-based music synthesis~\cite{diffrhythm} have enabled impressive progress in music captioning, style transfer, and conditional generation. However, existing systems still face significant limitations. Many music models rely on proprietary datasets or closed-source pipelines, which limits reproducibility and downstream research. Others provide only coarse control over musical attributes, lack robust alignment between textual descriptions and acoustic realizations, or struggle to maintain long-range musical coherence beyond short segments~\cite{musicgen}. Furthermore, end-to-end controllable song generation jointly guided by style descriptions, lyrics, and reference audio remains an open challenge.

To address these limitations, we introduce a family of open-source Music Foundation Models designed to unify music understanding, alignment, and controllable generation within a single extensible ecosystem. Our framework integrates four key components:

(1) \textbf{HeartCLAP}: an audio-text alignment model that learns a shared embedding space for music semantics, enabling accurate music tagging and cross-modal retrieval, and serving as a foundation for downstream generative tasks.

(2) \textbf{HeartTranscriptor}: a robust lyric recognition model tailored to complex musical signals, providing accurate transcription of lyrical content.

(3) \textbf{HeartCodec}: a low-frame-rate (12.5~Hz), high-fidelity music codec that captures both long-range structure and acoustic details. Its compact discrete representation enables high-quality reconstruction and efficient autoregressive modeling.

(4) \textbf{HeartMuLa}: a multi-condition song generator that accepts flexible user inputs, including style descriptions, detailed lyrics, and reference audio, while offering fine-grained control over musical attributes such as genre, mood, rhythm, and expressive variations.

Our song generation model supports long-form music creation of up to six minutes, maintaining both structural coherence and expressive diversity over extended durations. In addition, it provides two specialized modes: (1) a \emph{short-music generation} mode suitable for background music in short videos; and (2) a \emph{fine-grained style control} mode that enables creators to control the style of different song parts (e.g., intro, verse, chorus) using natural language prompts.

Beyond the capabilities of individual models, the open-source nature of our ecosystem enables reproducibility, extensibility, and broad community adoption. By releasing model weights and evaluation protocols, we aim to provide a comprehensive foundation for future advancements in music intelligence.

In summary, our contributions are as follows:
\begin{enumerate}
    \item We introduce a unified suite of open-source Music Foundation Models covering music--text alignment, music tokenization, lyric recognition, and controllable song generation.
    \item We propose a novel music codec tokenizer that achieves high expressiveness at a low frame rate, enabling efficient and scalable modeling of long musical sequences.
    \item We present a song generation framework with fine-grained musical control, supporting high-quality long-form generation of up to 6 minutes. 
\end{enumerate}

\section{HeartCodec}
\begin{figure}
        \centering
        \includegraphics[width=0.99\linewidth]{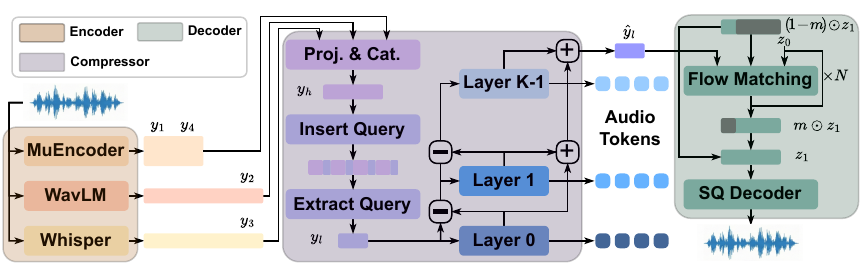}
        \caption{An illustration of our proposed \textbf{HeartCodec}. Left, middle, right are semantic-rich encoder, ultra-low frame rate compressor and high-fidelity reconstruction decoder, respectively.}
        \label{fig:heartcodec}
    \end{figure}

In this section, we first present the architecture of HeartCodec in \ref{HeartCodec_Architecture}, detailing its semantic-rich encoder, ultra-low-frame-rate compressor, and high-fidelity reconstruction decoder (see Fig. \ref{fig:heartcodec}). We then describe the training details in \ref{HeartCodec_Training}, including optimization objectives for different training stages and implementation specifics. Finally, \ref{HeartCodec_Experiment} reports our experimental results, covering comparisons with baselines and ablation studies on design choices.

\subsection{Architecture}
\label{HeartCodec_Architecture}
    HeartCodec compresses raw waveforms into discrete tokens through three parts, each contributing one of our core advantages: semantic-rich encoding, ultra-low-frame-rate tokenization, and high-fidelity reconstruction. We first employ pretrained audio encoders, including Whisper \cite{whisper}, WavLM \cite{wavlm}, MuEncoder \cite{mucodec_muencoder}, to produce semantically rich representations. Note that we use our training data to fine-tune the MuEncoder with the BEST-RQ \cite{best-rq,musicfm} loss. These representations are quantized via a query-based quantization strategy \cite{almtokenizer} to yield ultra-low-frame-rate discrete codes. Finally, a flow-matching-based decoder \cite{rectifiedflow_reflow} reconstructs the waveform from the quantized embeddings, achieving high-quality synthesis. 
    
\paragraph{Semantic-Rich Encoder}
The inherent complexity of music has motivated prior work to demonstrate that leveraging representations across multiple abstraction levels, rather than relying solely on low-level acoustic features, leads to stronger downstream modeling performance \cite{yue,levo,mucodec_muencoder,duotok}. Guided by this insight, we adopt a multi-encoder strategy and extract complementary representations from pretrained Whisper \cite{whisper}, WavLM \cite{wavlm}, and fine-tuned Music Encoder models (MuEncoder), covering phonetic, music semantic, and acoustic levels. 
Specifically, we extract eleventh-layer features as music semantic representations from MuEncoder, capturing high-level musical attributes such as timbre, phrasing, and melodic structure, while second-layer features are used as acoustic representations encoding fine-grained timbral and spectral details. To model phonetic-level cues related to vocal articulation and pronunciation, we use WavLM features obtained by averaging the outputs of layers 6 to 9, together with embeddings from the Whisper encoder.
Formally, given an input stereo waveform \( x \in \mathbb{R}^{T f_a \times C_a} \) with duration \(T\), sample rate \(f_a = ~\mathrm{48kHz}\) and channel \(C_a = 2\), we denote by \(y_i \in \mathbb{R}^{T f_i \times C_i}\) the feature sequence extracted from the \(i\)-th pretrained module, where \(f_i\) and \(C_i\) represent the corresponding frame rate and channel dimension. In particular, \(y_1\), \(y_2\), \(y_3\), and \(y_4\) correspond to the MuEncoder semantic features, WavLM phonetic features, Whisper embeddings, and MuEncoder acoustic features, respectively.
Together, this set of representations \(\{y_i\}_{i=1}^4\) forms a coherent multi-level description that jointly captures phonetic, music semantic, and acoustic information, providing a rich and structured foundation for downstream modeling.

\paragraph{Ultra-Low Frame Rate Compressor}
\label{HeartCodec_Tokenizer}
The frame rate of audio tokenizer directly determines sequence length; however, recent LM-based music generation systems \cite{musicgen,yue,levo} still rely on audio codecs operating at relatively high frame rates (25--50~Hz), which limits scalability. In contrast, our tokenizer operates at a substantially lower frame rate of 12.5~Hz while jointly fusing multi-level representations. Concretely, for each multi-level feature sequence \(y_i\), we independently resample it to a common frame rate \(f_h = 25~\mathrm{Hz}\). The temporally aligned features are fused into a single representation via channel-wise concatenation followed by a linear projection to dimension \(C\), yielding \(y_h \in \mathbb{R}^{T f_h \times C}.\). We then apply a query-based quantization strategy \cite{almtokenizer} on \(y_h\). After every two consecutive frames in \(y_h\), we insert a learnable query token, and process the resulting sequence with a Transformer encoder. The output embedding at each query-token position is retained as a learned summary of the two consecutive frames it follows, while all non-query frame embeddings are discarded. This reduces the frame rate to \(f_l = 12.5~\mathrm{Hz}\) and produces a downsampled representation \(y_l \in \mathbb{R}^{T f_l \times C}\), which is then discretized using a residual vector quantization (RVQ) module \cite{soundstream_rvq} with \(K = 8\) codebooks of vocabulary size \(V = 8192\), producing discrete indices \(A \in [V]^{T f_l \times K}\), which serve as the audio token sequence for downstream language-model-based generation. The corresponding quantized representation is obtained by codebook lookup and summation across \(K\) codebooks, yielding
\(\hat{y}_l \in \mathbb{R}^{T f_l \times C}\).
We employ the standard RVQ commitment loss, averaged over frames
\begin{equation}
\label{l_commit}
\mathcal{L}_{\mathrm{commit}} = \frac{1}{T f_l} \sum_{t=1}^{T f_l} 
\big\| \mathrm{sg}(y_{l,t}) - \hat{y}_{l,t} \big\|^2 .
\end{equation}
To preserve the rich semantic information in the discretized representation, we employ upsamplers \(\{U_i\}_{i=1}^4\) to align the quantized features \(\hat{y}_l\) with the reference features \(\{y_i\}_{i=1}^4\), yielding \(U_i(\hat{y}_l) \in \mathbb{R}^{T f_i \times C_i}\). Alignment is encouraged by minimizing the feature alignment loss:
\begin{equation}
\label{l_align}
\mathcal{L}_{\mathrm{align}}^{(i)} =
- \frac{1}{T f_i} \sum_{t=1}^{T f_i}
\log \text{sigmoid} \!\left(
\frac{{U_i(\hat{y}_l)}_{t}^\top y_{i,t}}
{\|{U_i(\hat{y}_l)}_{t}\| \, \|y_{i,t}\|}
\right).
\end{equation}
In practice, applying this loss to the MuEncoder semantic features and WavLM phonetic features (\(i=1,2\)) yields the best results.

\paragraph{High-Fidelity Reconstruction Decoder}

Directly reconstructing waveforms from discretized representations typically results in lower fidelity than their continuous counterparts, due to quantization-induced information loss. Following recent hybrid approaches \cite{mucodec_muencoder,almtokenizer,semanticodec,duotok}, we reconstruct waveforms by first mapping discrete representations into a continuous latent space using a generative model, followed by waveform reconstruction using the corresponding decoder. Specifically, we employ a pretrained continuous audio tokenizer \( G \) and extract continuous latents \( z = G_{\mathrm{Enc}}(x) \) as reconstruction targets. We experiment with several continuous audio tokenizer \cite{mucodec_muencoder,levo,simplespeech_sqcodec}, and select a 25 Hz SQ-Codec \cite{simplespeech_sqcodec} based on our ablation results (See Sec \ref{HeartCodec_Experiment}). The latent distribution is modeled using flow matching \cite{rectifiedflow_reflow}, where a vector field \( v_{\theta} \) transforms Gaussian noise \( z_0 \sim \mathcal{N}(0, I) \) into the target latent \( z_1 = z \), conditioned on low-frame-rate discretized features \( \hat{y}_l \). To enable seamless reconstruction across audio clips, we additionally condition the model on partial clean latents. Following \cite{mehta2024matcha,voicebox}, we apply a binary mask \( m \) and provide the unmasked latent \( (1 - m) \odot z_1 \) as input, while training the model to predict the masked portion \( m \odot z_1 \). Together, the training objective becomes:
\begin{equation}
\label{l_fm}
    \mathcal{L}_{\mathrm{fm}} =
 \mathop{\mathbb{E}}_{x, t, z_0}[\| v_{\theta}(z_t, \hat{y}_l, (1 - m) \odot z_1) - (m \odot(z_1 - z_0)) \|^2],
\end{equation}
where $z_t = tz_1 + (1-t)z_0$ is the noisy sample derived following schedule $t \sim U(0,1)$. We utilize a Diffusion Transformer \cite{dit} backbone based on LLaMA architecture \cite{llama3} for predicting $v_{\theta}$. All inputs $z_t, \hat{y}_l, (1 - m) \odot z_1$ are concated along feature channel. To further improve sampling efficiency, we adopt Reflow distillation \cite{rectifiedflow_reflow}, reducing the number of sampling steps from 50 to 10. The distillation objective is defined as
\begin{equation}
\label{l_distill}
\mathcal{L}_{\mathrm{distill}} =
\mathop{\mathbb{E}}\limits_{x, t, z_0}
\Big[
\big\|
v_{\theta}\big(z_t^{\theta^-}, \hat{y}_l, (1 - m) \odot z_1^{\theta^-}\big)
-
m \odot (z_1^{\theta^-} - z_0^{\theta^-})
\big\|^2
\Big],
\end{equation}
where \( \theta^- \) denotes a frozen copy of the trained vector field.
The latent trajectories \( z_t^{\theta^-} \) are obtained by integrating the frozen vector field,
\( z_t^{\theta^-} = z_0 + \int_{0}^{t} v_{\theta^-}(z_s, s)\, ds \). After distillation, the latent distribution induced by \( z_1^{\theta} \) still deviates from the ground-truth \( z_1 = G_{\mathrm{Enc}}(x) \), where \( x \sim p_{\mathrm{data}} \).
To bridge this gap, we further fine-tune the SQ-Codec decoder, adapting it to the distilled latent distribution by optimizing waveform reconstruction from \( \tilde{x} = G_{\mathrm{Dec}}(z_1^{\theta}) \), yielding the following objective:
\begin{equation}
\label{l_sq}
\mathcal{L}_{\mathrm{sq}} =
\mathcal{L}_{\mathrm{rec}}(\tilde{x}, x)
+
\lambda_{\mathrm{adv}} \, \mathcal{L}_{\mathrm{adv}}(\tilde{x}, x),
\end{equation}
where reconstruction loss \( \mathcal{L}_{\mathrm{rec}} \) consists of
time-domain and frequency-domain terms, including an \( \ell_1 \) loss between
the reconstructed waveform \( \tilde{x} \) and the ground-truth waveform \( x \),
as well as an MSE loss computed on the STFT spectrogram.
The adversarial loss \( \mathcal{L}_{\mathrm{adv}} \) is computed based on the discriminator outputs.

\subsection{Training Details}
\label{HeartCodec_Training}

Our training pipeline is conducted in three stages. 
We begin by pretraining and fine-tuning \textbf{HeartCodec} from scratch, which establishes its fundamental modeling and reconstruction capabilities. 
Next, we perform \textbf{ReFlow distillation} to accelerate inference while maintaining generation quality. 
Finally, we fine-tune \textbf{SQ-Codec} on top of the distilled model to further improve reconstruction fidelity. The parameters of MuEncoder, WavLM and Whisper modules are kept frozen during all stages.

\paragraph{Pretraining and Finetuning}
The overall loss is formulated as a weighted sum of the terms in
Eqs.~\ref{l_fm}, \ref{l_commit}, and \ref{l_align}, yielding
\begin{equation}
\mathcal{L}_{1} =
\lambda_{\mathrm{fm}} \, \mathcal{L}_{\mathrm{fm}}
+ \lambda_{\mathrm{commit}} \, \mathcal{L}_{\mathrm{commit}}
+ \lambda_{\mathrm{sem}} \, \mathcal{L}_{\mathrm{align}}^{(1)}
+ \lambda_{\mathrm{pho}} \, \mathcal{L}_{\mathrm{align}}^{(2)} .
\end{equation}
We set $\lambda_{\mathrm{fm}} = 1$, $\lambda_{\mathrm{commit}} = 1$, $\lambda_{\mathrm{sem}} = 0.1$, and $\lambda_{\mathrm{pho}} = 0.1$.
The last two terms correspond to feature alignment losses regarding semantic features and phonetic (WavLM) features, respectively. We train our model on an internal dataset consisting of about 600,000 songs.
During pretraining, we use a segment duration of $T = 20.48\,\mathrm{s}$, while a longer duration of $T = 29.76\,\mathrm{s}$ is adopted for fine-tuning.
All experiments are conducted on $8$ NVIDIA A100 GPUs with a global batch size of 160.
The model is trained for 15 epochs.
We use the AdamW optimizer with a base learning rate of $1\times10^{-4}$, together with a cosine learning rate scheduler that includes a warm-up period of the first $3\%$ steps. We use HeartCodec (Pt.) and HeartCodec (Pt. \& Ft.) to denote the model after pretraining only, and the model after pretraining followed by finetuning, respectively.

\paragraph{Reflow Distillation}
We perform reflow distillation on top of HeartCodec (Pt. \& Ft.). During the ReFlow distillation stage, we curated a collection of 50,000 high-quality segments, each with a duration of $T = 29.76$ s. Using the fine-tuned HeartCodec, we extracted triplets $(e_l, z_0^{\theta^-}, z_1^{\theta^-})$ consisting of the code, noise, and the model-predicted latent. Subsequently, we activated only the flow matching module and trained it on these 50,000 triplets according to Eq. \eqref{l_distill}. The training was conducted on 8 NVIDIA A100 GPUs for 2 epochs. We employed the AdamW optimizer with a learning rate of $5 \times 10^{-6}$. We use HeartCodec (Reflow) to denote the model after reflow distillation.

\paragraph{Finetuning SQ-Codec} Given the parameters of HeartCodec (Reflow), we further finetune SQ-Codec. In this stage, we freeze all parameters of HeartCodec except for the final SQ-Codec decoder.
The model is optimized using the loss $\mathcal{L}_{\mathrm{sq}}$ defined in Eq.~\ref{l_sq}.
Training is performed on a filtered high-quality subset of approximately 20k samples, selected using objective metrics from AudioBox Aesthetics \cite{tjandra2025metaaudioboxaestheticsunified} and SongEval \cite{yao2025songevalbenchmarkdatasetsong}. The model is trained for $3$ epochs on $4$ NVIDIA A100 GPUs. Both the generator and discriminator are optimized with the AdamW optimizer, using a learning rate of $2\times10^{-6}$ and an exponential learning rate scheduler with decay factor $\gamma=0.999$. The model after this stage is denoted as HeartCodec (SQ Ft.)

\subsection{Experiment Results}
\label{HeartCodec_Experiment}

\subsubsection{Experimental Setup}
\label{sec:exp_setup}
We conduct objective and subjective studies to evaluate HeartCodec comprehensively.

\textbf{Objective Metrics.} 
We assess performance across four critical dimensions: 
\begin{enumerate}
    \item \textbf{Vocal fidelity}, measured by Short-Time Objective Intelligibility (STOI), Perceptual Evaluation of Speech Quality (PESQ), Speaker Similarity (SPK\_SIM) computed by ECAPA-TDNN model \cite{desplanques2020ecapa}, and Word Error Rate (WER) whose transcript is recognized by our HeartTranscriptor; 
    \item \textbf{Overall reconstruction quality}, evaluated using Virtual Speech Quality Objective Listener (VISQOL), Fréchet Audio Distance (FAD), and Fréchet Distance (FD); 
    \item \textbf{Aesthetic metrics}, including Content Evaluation (CE), Content Understanding (CU), and Perceptual Quality (PQ) from AudioBox \cite{tjandra2025metaaudioboxaestheticsunified};
    \item \textbf{Musical Quality}, including Coherence (Coh.), Musicality (Mus.), Memorability (Mem.), Clarity (Cla.), and Naturalness (Nat.). These metrics are computed by SongEval \cite{yao2025songevalbenchmarkdatasetsong} and used in Sec. \ref{sec:heartcodec_ablation_training_stages} only.
    \item \textbf{Style Adherence}, measured by Tag Similarity (Tag-Sim.) defined as the cosine similarity between the embeddings of the generated audio and the prompt style tags, extracted via the MuQ-MuLan model \cite{zhu2025muq}. This metric is used in Sec. \ref{sec:heartcodec_ablation_training_stages} only.
    \item \textbf{Computational efficiency}, quantified by the Real-Time Factor (RTF), which measures the ratio of the processing time to the duration of the generated audio.
\end{enumerate}

\textbf{Subjective Metrics.} 
We conducted a blind listening test with five expert listeners with musical backgrounds. Evaluators rated each sample on a scale of 1 to 5 across the following aspects:
\begin{enumerate}
    \item \textbf{Reconstruction Similarity}, including Vocal Similarity (VS), Accompaniment Similarity (AS), and Mixing Similarity (MS), assessing how closely the components match the reference;
    \item \textbf{Acoustic Naturalness}, comprising Vocal Naturalness (VN) and Melody Naturalness (MN), focusing on the absence of artifacts and the fluidity of the melody;
    \item \textbf{Perceptual Fidelity}, involving Detail Retention (DR) and Overall Preference (Pref.), which reflect the model's ability to preserve nuanced textures and the general listening satisfaction.
\end{enumerate}

\textbf{Implementation Details.} 
The Flow Matching module in HeartCodec is configured with approximately 1.5B parameters. During the inference stage, the classifier-free guidance (CFG) scale is set to 1.25. For the sampling process, we employ 50 steps for the Pretrain and Finetune stages. To optimize inference efficiency, the sampling steps are reduced to 10 for the Reflow distillation and SQ-Finetune stages.

\subsubsection{Reconstruction Quality}
We conducted a comparative analysis against several representative music codecs, including SemantiCodec \cite{semanticodec}, XCodec \cite{xcodec}, MuCodec \cite{mucodec_muencoder}, and LeVo \cite{levo}. The results of the objective evaluation are summarized in Table \ref{tab:codec_comparison_recon}.

\begin{table}[htbp]
\centering
\caption{Comparative Evaluation between Existing Codec Models.}
\label{tab:codec_comparison_recon}
\setlength{\tabcolsep}{2.5pt} 
\resizebox{\textwidth}{!}{ 
\begin{tabular}{lccccccccccccc}
\toprule
\multicolumn{1}{c}{Model} & CodeBook & \begin{tabular}[c]{@{}c@{}}Framerate\end{tabular} & \begin{tabular}[c]{@{}c@{}}Bitrate\\ (kbps)\end{tabular} & VISQOL $\uparrow$ & FAD $\downarrow$ & FD $\downarrow$ & STOI $\uparrow$ & \begin{tabular}[c]{@{}c@{}}PESQ $\uparrow$\\ (WB/NB)\end{tabular} & SPK\_SIM $\uparrow$ & \begin{tabular}[c]{@{}c@{}}WER $\downarrow$\end{tabular} & CE $\uparrow$ & CU $\uparrow$ & PQ $\uparrow$ \\
\midrule
Ground Truth & - & - & - & - & - & - & - & - & - & 0.14 & 7.09 & 7.40 & 7.62 \\
\midrule
\multirow{2}{*}{SemantiCodec} & 1 x 32768 & 25 & 0.375 & 2.24 & 2.32 & 22.38 & 0.40 & 1.14/1.44 & 0.79 & 0.91 & 6.94 & 7.11 & 7.33 \\
 & 1 x 16384 & 100 & 1.40 & 2.28 & 1.92 & 16.90 & 0.53 & 1.30/1.82 & 0.89 & 0.52 & 6.91 & 7.12 & 7.31 \\ \midrule
\multirow{4}{*}{XCodec} & 1 x 1024 & 50 & 0.50 & 2.23 & 1.88 & 24.51 & 0.57 & 1.27/1.68 & 0.75 & 0.73 & 6.84 & 7.26 & 7.38 \\
 & 2 x 1024 & 50 & 1.00 & 2.28 & 1.20 & 18.30 & 0.64 & 1.43/1.95 & 0.84 & 0.52 & 6.86 & 7.21 & 7.41 \\
 & 4 x 1024 & 50 & 2.00 & 2.32 & 0.88 & 16.08 & 0.70 & 1.63/2.28 & 0.88 & 0.34 & 6.91 & 7.20 & 7.47 \\
 & 8 x 1024 & 50 & 4.00 & 2.35 & 0.70 & 14.78 & \textbf{0.74} & \textbf{1.87/2.62} & \textbf{0.91} & 0.27 & 6.91 & 7.22 & 7.50 \\ \midrule
MuCodec & 1 x 16384 & 25 & 0.35 & 3.07 & 1.02 & 14.73 & 0.45 & 1.12/1.36 & 0.76 & 0.54 & 6.99 & 7.26 & 7.59 \\ \midrule
LeVo (Mixed) & 1 x 16384 & 25 & 0.35 & 3.24 & 1.03 & 19.44 & 0.49 & 1.14/1.44 & 0.79 & 0.47 & 7.09 & 7.33 & 7.63 \\ 
LeVo (Dual-Track) & 2 x 16384 & 25 & 0.70 & 3.26 & 1.45 & 19.96 & 0.56 & 1.21/1.61 & 0.82 & 0.35 & \textbf{7.20} & \textbf{7.54} & \textbf{7.88} \\ \midrule
HeartCodec (Pt.) & 8 x 8192 & 12.5 & 1.30 & 3.57 & 0.52 & 12.33 & 0.64 & 1.45/2.02 & 0.90 & \textbf{0.26} & 7.06 & 7.26 & 7.48 \\
HeartCodec (Pt. \& Ft.) & 8 x 8192 & 12.5 & 1.30 & 3.57 & 0.61 & 12.20 & 0.64 & 1.46/2.04 & 0.90 & 0.27 & 7.06 & 7.27 & 7.49 \\
HeartCodec (Reflow) & 8 x 8192 & 12.5 & 1.30 & 3.61 & 0.46 & 11.65 & 0.64 & 1.45/2.03 & 0.90 & 0.27 & 7.05 & 7.29 & 7.49 \\
HeartCodec (SQ Ft.) & 8 x 8192 & 12.5 & 1.30 & \textbf{3.72} & \textbf{0.27} & \textbf{11.06} & 0.66 & 1.52/2.10 & 0.90 & \textbf{0.26} & 7.05 & 7.36 & 7.57 \\ \bottomrule
\end{tabular}
}
\end{table}

Experimental results demonstrate that Reflow distillation using high-quality data yields marginal improvements in overall reconstruction metrics (VISQOL, FAD, and FD). Subsequent SQ-Finetune leads to significant gains across all indicators, establishing HeartCodec as the state-of-the-art music tokenizer. Notably, HeartCodec exhibits a decisive advantage in global reconstruction quality: its VISQOL substantially outperforms all baselines, while its FAD and FD are the lowest recorded. This underscores its superior capability in aligning both time-domain distributions and frequency-domain features with the original audio.

Regarding vocal fidelity, HeartCodec remains in the top tier with STOI of 0.66 and PESQ of 1.52/2.10. Moreover, it demonstrates exceptional semantic and identity preservation, achieving the lowest WER and a competitive SPK\_SIM comparable to the high-bitrate XCodec. While XCodec achieves a higher STOI leveraging its higher bitrate and framerate, its integrated audio quality metrics like FAD and VISQOL still lag behind HeartCodec. In terms of aesthetics, the SQ-Finetune strategy allows HeartCodec to outperform most models in CE, CU, and PQ, trailing only slightly behind the dual-track architecture of LeVo.

\begin{table}[htbp]
\centering
\caption{Comparative Evaluation of HeartCodec Under Varying Guidance Scales.}
\label{tab:codec_comparison_recon_cfg_scale}
\setlength{\tabcolsep}{2.5pt} 
\resizebox{\textwidth}{!}{ 
\begin{tabular}{lccccccccccc}
\toprule
\multicolumn{1}{c}{Model} & Guidance Scale & VISQOL $\uparrow$ & FAD $\downarrow$ & FD $\downarrow$ & STOI $\uparrow$ & \begin{tabular}[c]{@{}c@{}}PESQ $\uparrow$\\ (WB/NB)\end{tabular} & SPK\_SIM $\uparrow$ & \begin{tabular}[c]{@{}c@{}}WER $\downarrow$\end{tabular} & CE $\uparrow$ & CU $\uparrow$ & PQ $\uparrow$ \\ \midrule
HeartCodec (Pt.) & 1.25 & 3.57 & 0.52 & 12.33 & 0.64 & 1.45/2.02 & 0.90 & 0.26 & 7.06 & 7.26 & 7.48 \\
HeartCodec (Pt. \& Ft.) & 1.25 & 3.57 & 0.61 & 12.20 & 0.64 & 1.46/2.04 & 0.90 & 0.27 & 7.06 & 7.27 & 7.49 \\
HeartCodec (Pt.) & 1.5 & 3.51 & 1.21 & 17.65 & 0.68 & 1.49/2.13 & 0.89 & 0.24 & 6.98 & 7.11 & 7.39 \\
HeartCodec (Pt. \& Ft.) & 1.5 & 3.61 & 0.61 & 14.02 & 0.68 & 1.56/2.18 & 0.90 & 0.24 & 7.07 & 7.35 & 7.62 \\\bottomrule
\end{tabular}
}
\end{table}

We further evaluate the performance of the fine-tuning stage across different guidance scales in Table \ref{tab:codec_comparison_recon_cfg_scale}. While the objective metrics show marginal improvement at a guidance scale of 1.25, the performance gains at a guidance scale of 1.5 are substantial, underscoring the effectiveness of our fine-tuning approach. However, despite the superior objective scores at guidance scale 1.5, subjective assessments reveal that a guidance scale of 1.25 yields a more natural and balanced auditory experience, with vocals and accompaniment sounding smoother and less harsh. Consequently, we select 1.25 as our default guidance scale.

As the ideal tokenizer for the HeartMuLa, HeartCodec's strength lies not only in high reconstruction quality and rich semantic capture but also in its breakthrough regarding extremely low frame rate in music tokenizers. Given HeartMuLa's Global-Local architecture, where a lightweight Local Transformer predicts a multi-layer RVQ, the computational latency added by increasing the RVQ layers is significantly lower than the linear growth caused by increasing the frame rate. HeartCodec’s low frame rate and multi-layer RVQ design enhance the scalability of HeartMuLa, providing a robust foundation for high-fidelity, streaming music generation.

\subsubsection{Ablation Study of the Latent Space}
We utilize the bottleneck features from three distinct models, Mel VAE, 1D VAE, and SQ-Codec, as the target latents for the Flow Matching model. Specifically, 1D VAE and SQ-Codec reconstruct waveforms directly from their respective latents, whereas Mel VAE first recovers the Mel-spectrogram, followed by waveform synthesis via HiFi-GAN \cite{kong2020hifi}. We evaluate the reconstruction fidelity and computational efficiency of HeartCodec under these different objective latents upon completion of pre-training. This analysis aims to explore the impact of various latent representations on music reconstruction, with experimental results summarized in Table \ref{tab:codec_comparison_recon_latent_feature}.

\begin{table}[htbp]
\centering
\caption{Comparative Evaluation of HeartCodec Under Varying Latent Features.}
\label{tab:codec_comparison_recon_latent_feature}
\setlength{\tabcolsep}{2.5pt} 
\resizebox{\textwidth}{!}{ 
\begin{tabular}{lcccccccc}
\toprule
\multicolumn{1}{c}{Model} & VISQOL $\uparrow$ & FAD $\downarrow$ & FD $\downarrow$ & STOI $\uparrow$ & \begin{tabular}[c]{@{}c@{}}PESQ $\uparrow$\\ (WB/NB)\end{tabular} & SPK\_SIM $\uparrow$ & WER $\downarrow$ & RTF $\downarrow$ \\ \midrule
HeartCodec (Mel VAE) & 3.43 & \textbf{0.51} & \textbf{8.15} & 0.54 & 1.27/1.74 & 0.88 & 0.34 & 0.397 \\
HeartCodec (1D VAE) & \textbf{3.57} & 0.84 & 12.95 & 0.63 & \textbf{1.46}/\textbf{2.06} & 0.89 & 0.32 & \textbf{0.116} \\
HeartCodec (SQ-Codec) & \textbf{3.57} & 0.52 & 12.33 & \textbf{0.64} & 1.45/2.02 & \textbf{0.90} & \textbf{0.26} & 0.121 \\
\bottomrule
\end{tabular}
}
\end{table}

Compared to 1D VAE, SQ-Codec achieves significantly superior scores in distribution-based metrics, specifically FAD and FD, while maintaining comparable performance across signal-level reconstruction metrics (VISQOL, STOI, and PESQ) and computational efficiency (RTF). Conversely, although Mel VAE exhibits lower FAD and FD, it is markedly inferior to SQ-Codec and 1D VAE in all other objective reconstruction indicators. Furthermore, the prohibitive RTF of Mel VAE (0.397) poses a substantial bottleneck for real-time streaming music generation.

\begin{table}[htbp]
\centering
\caption{Subjective Evaluation of HeartCodec Under Different Latent Features.}
\label{tab:subjective_comparison}
\setlength{\tabcolsep}{3.5pt} 
\resizebox{\textwidth}{!}{ 
\begin{tabular}{lccccccc}
\toprule
\multicolumn{1}{c}{Model} & VS $\uparrow$ & AS $\uparrow$ & MS $\uparrow$ & VN $\uparrow$ & MN $\uparrow$ & DR $\uparrow$ & Pref. $\uparrow$ \\ \midrule
HeartCodec (Mel VAE)   & 3.58 & 3.68 & 2.75 & 3.43 & 4.00 & 3.23 & 3.33 \\
HeartCodec (1D VAE)    & 3.30 & 3.35 & 2.90 & 3.21 & 3.70 & 3.00 & 2.90 \\
HeartCodec (1D SQ)     & \textbf{3.90} & \textbf{4.00} & \textbf{3.33} & \textbf{3.79} & \textbf{4.25} & \textbf{3.78} & \textbf{3.85} \\ 
\bottomrule
\end{tabular}
}
\end{table}

To further bridge the gap between objective metrics and human auditory perception, we assess the models' subjective performance in Table \ref{tab:subjective_comparison}. SQ-Codec outperforms its counterparts across all seven subjective dimensions. Notably, in terms of Vocal Similarity (VS) and Melody Naturalness (MN), SQ-Codec achieves scores of 3.90 and 4.25, respectively, significantly higher than 1D VAE. Taking all factors into consideration, we follow the setting of SimpleSpeech \cite{simplespeech_sqcodec,simplespeech2}, and designate the bottleneck features of SQ-Codec as the target latent for the Flow Matching module.

\subsubsection{Ablation Study of Training Stages}
\label{sec:heartcodec_ablation_training_stages}
To further assess the impact of reflow distillation and fine-tuning SQ-Codec on downstream music generation tasks, we conduct an additional set of experiments. Specifically, our music generation model, HeartMuLa (see Section~\ref{HeartMuLa_arch}), is trained to model token sequences produced by HeartCodec. During inference, we let HeartMuLa predict a sequence of tokens given lyrics and style tags as prompts, keep the predicted tokens fixed, and decode them using HeartCodec obtained at different training stages.

Specifically, we consider HeartCodec after pretraining and finetuning (Pt. \& Ft.), after reflow distillation (Reflow), and after SQ-Codec fine-tuning (SQ Ft.). Notably, during both reflow distillation and SQ-Codec fine-tuning, the encoder and compressor components of HeartCodec are kept frozen. Therefore, using the same token sequence as input across different decoders is well justified and allows for a controlled comparison.

After decoding, we evaluate the generated music using metrics including aesthetic metrics produced by AudioBox \cite{tjandra2025metaaudioboxaestheticsunified}, musical quality measured by SongEval \cite{yao2025songevalbenchmarkdatasetsong}, style alignment measured by tag similarity (Tag-Sim) and intelligibility of the vocal track measured by the phoneme error rate (PER). We report results on our  HeartBeats-Benchmark (English) dataset (see Sec. \ref{heartbeats_benchmark}), as summarized in Table~\ref{tab:ablation of training stages}. As shown in the results, compared to HeartCodec (Pt. \& Ft.), the model after reflow distillation achieves noticeable improvements in both aesthetic quality and tag similarity, albeit with a degradation in vocal intelligibility. After further fine-tuning the SQ-Codec , all evaluation metrics consistently improve, yielding the best overall performance among the compared methods. These results demonstrate that the two additional training stages, reflow distillation and SQ-Codec fine-tuning, consistently enhance the quality of downstream music generation.

\begin{table*}[htbp]
\centering
\caption{Objective Evaluation of the Effects of Training Stages on Downstream Music Generation.}
\label{tab:ablation of training stages}
\setlength{\tabcolsep}{3.5pt} 
\resizebox{\textwidth}{!}{ 
\begin{tabular}{lccccccccccc}
\toprule
\multirow{2}{*}{\textbf{Model}} & \multicolumn{3}{c}{\textbf{AudioBox} $\uparrow$} & \multicolumn{6}{c}{\textbf{SongEval} $\uparrow$} & \multicolumn{1}{c}{\textbf{Align} $\uparrow$} & \multicolumn{1}{c}{\textbf{Intel} $\downarrow$} \\
\cmidrule(lr){2-4} \cmidrule(lr){5-10} \cmidrule(lr){11-11} \cmidrule(lr){12-12}
 & CE & CU & PQ & Coh. & Mus. & Mem. & Cla. & Nat. & Avg. & Tag-Sim & PER \\ \midrule
HeartCodec (Pt. \& Ft.) & 7.37 & 7.78  & 8.05 & 4.43 & 4.24 & 4.43 & 4.30 & 4.13 & 4.31 & 0.2339 & 0.1092 \\
HeartCodec (Reflow) & 7.42 & 7.78  & 8.04 & 4.46 & 4.28 & 4.46 & 4.32 & 4.18 & 4.34 & 0.2486 & 0.1258 \\
HeartCodec (SQ Ft.) & \textbf{7.45} & 7.78  & \textbf{8.07} & \textbf{4.52} & \textbf{4.36} & \textbf{4.54} & \textbf{4.38} & \textbf{4.25} & \textbf{4.41} & \textbf{0.2499} & \textbf{0.1005} \\
\bottomrule
\end{tabular}
}
\end{table*}

\section{HeartMuLa}
\begin{figure}
        \centering
        \includegraphics[width=0.9\linewidth]{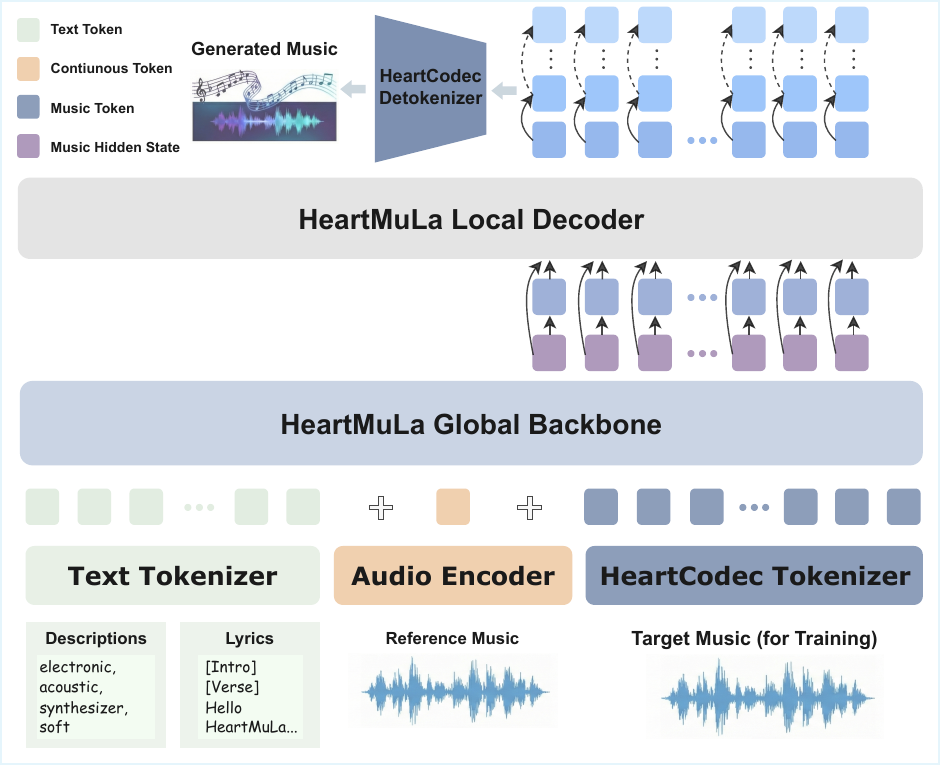}
        \caption{HeartMuLa Architecture}
        \label{fig:HeartMuLa}
\end{figure}

In this section, we introduce \textbf{HeartMuLa} (\textbf{Heart} \textbf{Mu}sic \textbf{La}nguage Model), a novel framework for music generation that operates on discrete audio tokens produced by our \textbf{HeartCodec}. HeartMuLa is designed to deliver long-form, high-fidelity, and controllable music synthesis. The section is organized as follows: Sec. \ref{HeartMuLa_arch} presents the hierarchical architecture, which enables efficient and high-fidelity generation; Sec. \ref{HeartMuLa_cond} details the integration of diverse conditions for precise control. The overall architecture of HeartMuLa is illustrated in Fig. \ref{fig:HeartMuLa}; Sec. \ref{HeartMuLa_training} describes the training strategy and implementation details; and Sec. \ref{HeartMuLa_eval} presents the evaluation of the model.

\subsection{Hierarchical Architecture}
\label{HeartMuLa_arch}
Following previous work \cite{yang2023uniaudio}, HeartMuLa employs a hierarchical factorization of the modeling process, initially capturing the coarse, long-range musical structure and subsequently incorporating fine-grained acoustic details. Specifically, we adopt our HeartCodec to tokenize audio into RVQ token sequence \( A = [a_0, a_1, \dots, a_{L-1}] \in [V]^{L \times K} \), where \( L = T f_a \) denotes the number of frames, and each frame \( a_l = [a_{l,0}, \dots, a_{l,K-1}] \) consists of \( K \) RVQ tokens. Let \( h_{l,k} \) denote the embedding of token \( a_{l,k} \), and define \( h_l = \sum_{k=0}^{K-1} h_{l,k} \) as the embedding of frame \( a_l \). The modeling process is decomposed into global and local stages: a global transformer \( \theta_{\text{glo}} \) which first models intra-frame dependencies by predicting \( a_{l,0} \) at each frame \( l \), conditioned on the preceding frame embeddings \( h_{<l} \), capturing the coarse semantic information encoded in the RVQ Layer 0 code, followed by a subsequent local transformer \( \theta_{\text{loc}} \) predicting the remaining tokens \( a_{l,k} \) within this frame, conditioned on both the hidden state of the global transformer \( \theta_{\text{glo}}(h_{<l}) \) and local token embeddings \( h_{l,<k} \). The overall probability is given by:

\begin{equation}
\label{Eq:uniaudio_eq}
    p(a_l \mid h_{<l}; \theta_{\text{glo}}, \theta_{\text{loc}}) = p(a_{l,0} \mid h_{<l}; \theta_{\text{glo}}) \left( \prod_{k=1}^{K-1} p(a_{l,k} \mid h_{l,<k}, \theta_{\text{glo}}(h_{<l}); \theta_{\text{loc}}) \right)
\end{equation}

This hierarchical architecture simultaneously delivers computational efficiency and high-fidelity generation capabilities. Enhanced Efficiency is achieved through sequence factorization; the large-scale global transformer is tasked solely with predicting the base tokens of layer 0 rather than the entire multi-layer hierarchy. This strategy significantly mitigates the computational overhead and modeling complexity associated with predicting multi-stream codebooks. Furthermore, the architecture leverages the global transformer to capture coarse-grained semantic and structural patterns across frames, while offloading the synthesis of intricate acoustic details to the local transformer. These components work in tandem to elevate the overall quality of the generated audio synergistically.

\subsection{Conditioning Mechanism}
\label{HeartMuLa_cond}

HeartMuLa uses lyrics, complemented by optional tags and reference audio as conditioning signals to enable precise control over the generation process.

\textbf{Lyrics} are annotated with structural markers such as \texttt{[intro]}, \texttt{[verse]}, and \texttt{[chorus]}, which guide the model in identifying and preserving the song’s structure. These markers are retained during tokenization by the Llama-3.2 tokenizer \cite{llama3}, yielding lyrics token sequence which is further embedded into $C_{\text{lyrics}}$.

\textbf{Tags} capture high-level musical attributes, each falling under a specific category (e.g., genre, instrument), with varying levels of influence on the music. To prioritize categories with greater impact, we empirically assign a selection probability to each category, ensuring that more influential tags, such as genre, are given higher probabilities compared to less impactful ones, such as topic. The specific categories and their selection probabilities are provided in Table \ref{tab:tag_probabilities}. These selected descriptions are then encapsulated within special tokens \texttt{<tag>} and \texttt{</tag>}, followed by tokenization via the Llama-3.2 tokenizer \cite{llama3} and embedding to form the tag sequence, represented as $C_{\text{tag}}$.

\textbf{Reference Audio} serves as a global stylistic cue. During training, we randomly sample a 10-second segment from the ground-truth music and employ pre-trained MuQ-MuLan \cite{zhu2025muq} embeddings \footnote{We do not want to release a model that can copy speaker's timbre, and we find that MuQ-MuLan does not include any speaker timbre information. Thus, we choose MuQ-Mulan to extract the style embedding.} to characterize its musical style. This conditioning signal, denoted as $C_{\mathrm{muq}}$, is discarded with a 50\% probability during training to facilitate unconditional modeling.

Together, $C_{\text{lyrics}}$, $C_{\text{tag}}$ and $C_{\text{muq}}$ form our overall condition sequence $C$, which is prepended before $h_0$
 and integrated as the prompt condition. This leads to a slight modification in the training objective compared to Eq. \ref{Eq:uniaudio_eq}, and will be detailed in Sec. \ref{HeartMuLa_objective}. Additionally, not all conditions are used during every training stage, as will be explained in Sec. \ref{HeartMuLa_training_stage}.

\begin{table}[htbp]
\centering
\caption{Selection Probabilities for Different Tag Categories}
\begin{tabular}{c|cccccccc}
\toprule
\textbf{Category} & Genre & Timbre & Gender & Mood & Instr. & Scene & Region & Topic \\ \midrule
\textbf{Prob.} & 0.95 & 0.5 & 0.375 & 0.325 & 0.25 & 0.2 & 0.125 & 0.1 \\ 
\bottomrule
\end{tabular}
\label{tab:tag_probabilities}
\end{table}

\subsection{Training}
\label{HeartMuLa_training}

\begin{figure}
        \centering
        \includegraphics[width=1.0\linewidth]{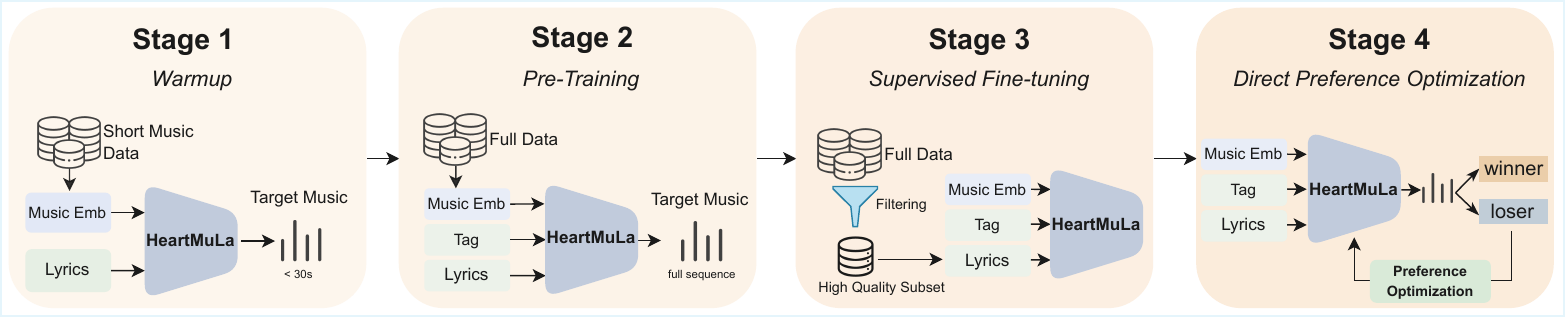}
        \caption{Four-Stage Progressive Training Paradigm}
        \label{fig:placeholder}
\end{figure}

We adopt a four-stage progressive training paradigm, as depicted in Fig. \ref{fig:placeholder}, which includes warm-up, pre-training, supervised fine-tuning, and reinforcement learning. We introduce each stage briefly in Sec. \ref{HeartMuLa_training_stage} and corresponding training objectives in Sec. \ref{HeartMuLa_objective}. Implementation details are provided in Sec. \ref{HeartMuLa_imp}.

\subsubsection{Four-Stage Progressive Training Paradigm}
\label{HeartMuLa_training_stage}

\textbf{Stage 1: Warmup.} In this stage, we train HeartMuLa on 30-second music segments containing lyrics. The input conditions $C = [C_{\text{muq}}, C_{\text{lyrics}}]$ include the lyrics and the reference audio. The core objective of this stage is to facilitate rapid parameter convergence and establish a preliminary mastery of local acoustic texture modeling laws through dense training on short contexts, laying a foundational acoustic capability for subsequent long-sequence generation. 

\textbf{Stage 2: Pretraining.} In this stage, HeartMuLa is trained on a large-scale dataset of full songs, using all three conditions $C = [C_{\text{tag}}, C_{\text{muq}}, C_{\text{lyrics}}]$. This stage aims to leverage massive data to force the model to learn long-range temporal dependencies and global musical structures under complete conditional inputs, thereby achieving precise adherence to lyrical content and effective control over the overall musical style.

\textbf{Stage 3: Supervised Finetuning.} In this stage, we use AudioBox \cite{tjandra2025metaaudioboxaestheticsunified} and SongEval \cite{yao2025songevalbenchmarkdatasetsong} to filter a high-quality subset from the full data. Then we fine-tune the model using all conditions $C = [C_{\text{tag}}, C_{\text{muq}}, C_{\text{lyrics}}]$. This stage aims to improve both the synthesis quality and the fine-grained structural control of the generated music via fine-tuning optimization.

\textbf{Stage 4: Direct Preference Optimization.} 
To further elevate the perceptual quality of the generated music, we use a reinforcement alignment stage utilizing Direct Preference Optimization (DPO) \cite{rafailov2023direct}. By leveraging preference data constructed with metrics including tag similarity, Phonemes Error Rate (PER), and AudioBox \cite{tjandra2025metaaudioboxaestheticsunified} scores, DPO enables the model to effectively distinguish and generate superior music. This stage is specifically designed to enhance multi-dimensional generation quality, resulting in significant improvements in vocal clarity, stylistic fidelity, and overall audio coherence.

\subsubsection{Optimization Objectives}
\label{HeartMuLa_objective}
\paragraph{Weighted CrossEntropy Loss} 
Based on the hierarchical modeling architecture of HeartMuLa, our optimization objective is decomposed into a global loss $\mathcal{L}_{0}$ targeting long-term semantic modeling and residual terms $\mathcal{L}_{k}, (k \in [1,K-1])$ focusing on acoustic detail reconstruction. The total loss function $\mathcal{L}_{\text{total}}$ is defined as follows:
\begin{equation}
\label{Eq. WCELoss}
\mathcal{L}_{\text{total}} = \lambda_0\mathcal{L}_0+\frac{1}{K-1}\sum_{k=1}^{K-1}\lambda_k\mathcal{L}_{k}
\end{equation}

where each term corresponds to the cross-entropy loss at the specific RVQ layer:
\begin{equation}
\mathcal{L}_{0} = -\frac{1}{L-1} \sum_{l=1}^{L-1} \log p(a_{l,0} \mid h_{<l}, C; \theta_{\text{glo}})
\end{equation}
\begin{equation}
\mathcal{L}_{k} = -\frac{1}{L-1} \sum_{l=1}^{L-1}  \log p(a_{l,k} \mid h_{l,<k}, \theta_{\text{glo}}(h_{<l}, C); \theta_{\text{loc}}), \forall k \in [1, K-1]
\end{equation}
Here, $\lambda_0$ is the weight for loss at layer 0, and \(\lambda_k\)  for the loss at the \(k\)-th residual layer. We assign more weight to layer 0 by averaging the residual terms, as layer 0 captures the coarse semantic information. The total loss function $\mathcal{L}_{\text{total}}$ integrates the prefix condition \(C\) into Eq. \ref{Eq:uniaudio_eq}, applying the negative log-likelihood followed by weighted summation.

During the Warmup and Pretraining stages, we aim for the model to learn the acoustic feature distribution across all layers in a balanced manner. Consequently, we set the global loss coefficient to $\lambda_0=1.0$ and assign uniform weights to all residual layers, i.e., $\lambda_k = 1.0, \forall k \in [1, K-1]$. During the Supervised Finetuning stage, we raise the weight of global loss to $\lambda_0=2.0$ to emphasize musical structure and semantics, while weights for the remaining terms are set to $\lambda_k = \frac{K-k}{10}$ to attenuate their influence.

\paragraph{Direct Preference Optimization Loss}

Traditional Reinforcement Learning (RL) methods for LLMs (e.g., PPO) ~\cite{schulman2017proximal} rely on an explicit reward model and online sampling, leading to substantial computational overhead and training instability. We instead adopt Direct Preference Optimization (DPO)~\cite{rafailov2023direct}, which casts the RL objective as supervised learning and directly optimizes the policy with preference pairs.

Given a dataset of preference pairs $\mathcal{D} = \{(C, A_{wn}, A_{ls})\}$, where $C$ denotes the prompt (including style description, structure lyrics and optional reference music), $A_{wn}$ represents the preferred (winning) audio sequence, and $A_{ls}$ represents the dispreferred (losing) audio sequence, DPO derives a closed-form solution to the constrained reward maximization problem. The resulting objective bypasses explicit reward modeling by expressing the optimal policy directly in terms of the log-probability ratio between the policy model $p_\theta$ and a reference model $p_{\text{ref}}$:
\begin{equation}
\label{eq dpo}
\mathcal{L}_{\text{DPO}}(\theta) = -\mathbb{E}_{(C, A_{wn}, A_{ls}) \sim \mathcal{D}} \left[ \log \sigma \left( \beta \cdot \Delta_\theta(C, A_{wn}, A_{ls}) \right) \right],
\end{equation}
where $\sigma(\cdot)$ denotes the sigmoid function, $\beta$ is a temperature hyperparameter controlling the deviation from the reference policy, and
\begin{equation}
\Delta_\theta(C, A_{wn}, A_{ls}) = \left( \log p_\theta(A_{wn} | C) - \log p_{\text{ref}}(A_{wn} | C) \right) - \left( \log p_\theta(A_{ls} | C) - \log p_{\text{ref}}(A_{ls} | C) \right).
\end{equation}

This formulation implicitly defines the reward as $r(C, A) = \beta \log \frac{p_\theta(A|C)}{p_{\text{ref}}(A|C)}$, enabling the model to learn from preference comparisons without requiring reward labels or online generation during training.

HeartMuLa employs a hierarchical modeling process where layer 0 code is modelled by global transformer and residual codes by local transformer. This design necessitates a decomposition of the log-probability. Similar to Eq. \ref{Eq. WCELoss}, the log-probability is factorized as follows:

\begin{equation}
\label{eq: log prob}
\log p_\theta(A | C) = \sum_{l=1}^{L} \log p(a_{l,0} \mid h_{<l}; \theta_{\text{glo}}) + \sum_{l=1}^{L} \sum_{k=1}^{K-1} \log p(a_{l,k} \mid h_{l,<k}, \theta_{\text{glo}}(h_{<l}); \theta_{\text{loc}}).
\end{equation}

Substituting Eq. \ref{eq: log prob} into Eq. \ref{eq dpo} yields the final DPO loss, which is naturally decomposed into two additive terms according to the hierarchical structure of HeartMuLa: one associated with the global semantic codes (layer-0) and another associated with the local residual acoustic codes (layers $1$ to $K-1$). Importantly, this decomposition implies that preference signals (rewards) can independently influence global semantic coherence and local acoustic details. As a result, the model can adjust its global semantic planning and local acoustic rendering in a disentangled manner, enabling more stable and targeted optimization compared to applying a single monolithic DPO objective over all tokens.

\subsubsection{Implementation Details}
\label{HeartMuLa_imp}

Built upon the Llama 3.2 architecture \cite{llama3}, HeartMuLa integrates a 3B-parameter global backbone with a 300M-parameter local decoder, optimized using the training strategy detailed in Section \ref{HeartMuLa_training}.

\noindent\textbf{Warmup Details.} We initiated training with a 10,000-hour subset randomly sampled from our 100,000-hour high-quality music corpus, segmented into 30-second clips. This stage ran on 8 NVIDIA A100 80GB GPUs for 5 epochs. We utilized the AdamW optimizer with a Cosine scheduler, setting the learning rate to $2e-4$ and the CFG dropout probability to 0.02 to support Classifier-Free Guidance.

\noindent\textbf{Pretraining Details.} Subsequently, we scaled the training to the full 100,000-hour dataset on 64 NVIDIA A100 GPUs. The model was trained for another 5 epochs. While maintaining the optimizer configuration and CFG dropout, we annealed the learning rate to $2e-5$ to ensure stability.

\noindent\textbf{SFT Details.} This stage employed a curated dataset of 15,000 hours of high-quality music. Training was executed on 8 NVIDIA A100 GPUs for 3 epochs, using the same learning rate $2e-5$ and hyperparameters as the pre-training stage.

\noindent\textbf{DPO Data Preparation.} 
To construct the preference data for DPO, we employed a divergent sampling strategy. For each prompt input, the model generated four candidate audio samples with varying quality. Based on these candidates, we curated three distinct preference datasets, each prioritizing different metrics to guide the optimization process. In all cases, a preference pair $(y_w, y_l)$ consists of a chosen winner sample and a loser sample. The specific criteria are as follows:

\begin{itemize}
    \item \textbf{Muq-similarity-based Set:} To enhance semantic alignment, the candidate with the highest Muq-similarity score was selected as $y_w$, and the one with the lowest score as $y_l$. We enforced a margin of $sim_w - sim_l > 0.12$ to ensure sufficient discriminability, and required $sim_w > 0.3$ to guarantee the fundamental quality of the positive sample.
    \item \textbf{PER-based Set:} Focusing on articulation accuracy, we selected the sample with the lowest Phoneme Error Rate (PER) as $y_w$ and the highest as $y_l$. A margin constraint of $|\text{PER}_w - \text{PER}_l| > 0.1$ was applied.
    \item \textbf{Audiobox \& SongEval-based Set:} To capture holistic audio quality, $y_w$ was identified as the sample achieving superior scores on both Audiobox \cite{tjandra2025metaaudioboxaestheticsunified}  and SongEval \cite{yao2025songevalbenchmarkdatasetsong} metrics, while $y_l$ performed worst on both. We filtered pairs to ensure a SongEval score margin $> 0.5$ and an Audiobox average score margin $> 0.8$.  
\end{itemize}

\noindent\textbf{Training Configuration.}
For the DPO training phase, we set the learning rate to $1 \times 10^{-7}$ and the KL penalty parameter $\beta$ to 0.1. The model was trained for 3 epochs on a cluster of 8 NVIDIA A100 GPUs.

\subsection{Evaluation}
\label{HeartMuLa_eval}

\subsubsection{Evaluation Setup}

We conduct both objective and subjective studies to comprehensively evaluate HeartMuLa.

\textbf{Objective Metrics.} To assess the performance of HeartMuLa, we employ a set of objective metrics focusing on three key aspects: Music Quality, Style Adherence, and Lyric Intelligibility. 

\begin{enumerate}
    \item \textbf{Musical Quality}: To assess the general audio quality and musical structure, we adopt two comprehensive metrics for generated music. \textbf{SongEval} \cite{yao2025songevalbenchmarkdatasetsong} evaluates structural and musical aspects, including Coherence (Coh.), Musicality (Mus.), Memorability (Mem.), Clarity (Cla.), and Naturalness (Nat.). \textbf{AudioBox} \cite{tjandra2025metaaudioboxaestheticsunified} focuses on audio quality, providing scores for Content Evaluation (CE), Content Understanding (CU), and Perceptual Quality (PQ).
    \item \textbf{Style Adherence}: To evaluate the controllability of our model, we utilize the \textbf{Tag Similarity}. This metric computes the cosine similarity between the embeddings of the generated audio and the target style tags, extracted via the MUQ-MuLan model \cite{zhu2025muq}.
    \item \textbf{Lyric Intelligibility}: To measure how clearly the model generates lyrics, we calculate the \textbf{Phoneme Error Rate (PER)}. Specifically, we extract vocals from the generated music using Demucs \cite{defossez2019demucs}, transcribe them with our HeartTranscriptor in Section \ref{heart_transcriptor}, and compute the phoneme-level discrepancy between the transcription and the reference lyrics.
\end{enumerate}

\textbf{Subjective Metrics.} To capture perceptual nuances that objective metrics might miss, we conducted a blind listening test. Human evaluators rated the generated samples on a 5-point Mean Opinion Score (MOS) scale across six distinct dimensions:

\begin{enumerate} \item \textbf{Musicality}: Focuses on the aesthetic appeal, specifically assessing the pleasantness of the melody and the depth of emotional expression. \item \textbf{Harmony}: Evaluates the acoustic cohesion, measuring how naturally the vocals blend with the accompaniment. \item \textbf{Structure}: Examines the organization of the song, checking for a clear musical progression and structural completeness. \item \textbf{Fidelity}: Measures the pure audio quality, specifically the clarity of sound and the absence of background noise or artifacts. \item \textbf{Creativity}: Assesses the novelty and inventiveness of the musical content. \item \textbf{Memorability}: Reflects the "catchiness" of the song, indicating how easily the melody stays in the listener's mind. \end{enumerate}

\textbf{Implementation Details.} During inference, we set the CFG scale to 1.5, temperature to 1.0, and top-k to 50. We use \textbf{HeartCodec} for detokenization. All evaluation experiments are conducted on NVIDIA A100 80GB GPUs.

\subsubsection{Objective Evaluation}

We evaluated representative open-source models including LeVo \cite{levo}, DiffRhythm 2 \cite{jiang2025diffrhythm}, YuE \cite{yue}, and ACE-Step \cite{gong2025ace} and closed-source models including Suno-v5, Suno-v4.5, Mureka-V7.6, Udio-v1.5, and MiniMax-Music-2.0.

To conduct a comprehensive objective evaluation, we utilize the multilingual HeartBeats Benchmark introduced in Section \ref{heartbeats_benchmark}. To ensure fair comparison, the input conditions are strictly limited to lyrics and textual style descriptions, as audio reference prompts are not supported by all baselines. Furthermore, models with inherent language constraints (e.g., LeVo \cite{levo}) are evaluated solely on their supported languages. The detailed objective results are reported separately by language: Table \ref{tab:heartmula_main_result_en} presents the performance on the English dataset, followed by Chinese in Table \ref{tab:heartmula_main_result_zh}, Japanese in Table \ref{tab:heartmula_main_result_jp}, Korean in Table \ref{tab:heartmula_main_result_ko}, and Spanish in Table \ref{tab:heartmula_main_result_es}.

\begin{table*}[htbp]
\centering
\caption{Objective evaluation results on the HeartBeats Benchmark (English).}
\label{tab:heartmula_main_result_en}
\setlength{\tabcolsep}{3.5pt} 
\resizebox{\textwidth}{!}{ 
\begin{tabular}{lcccccccccccc}
\toprule
\multirow{2}{*}{\textbf{Model}} & \multicolumn{4}{c}{\textbf{AudioBox} $\uparrow$} & \multicolumn{6}{c}{\textbf{SongEval} $\uparrow$} & \multicolumn{1}{c}{\textbf{Align} $\uparrow$} & \multicolumn{1}{c}{\textbf{Intel} $\downarrow$} \\
\cmidrule(lr){2-5} \cmidrule(lr){6-11} \cmidrule(lr){12-12} \cmidrule(lr){13-13}
 & CE & CU & PC & PQ & Coh. & Mus. & Mem. & Cla. & Nat. & Avg. & Tag-Sim & PER \\ \midrule
Suno-v5 & 7.65 & 7.83 & \textbf{6.46} & 8.21 & \textbf{4.62} & \textbf{4.51} & \textbf{4.63} & \textbf{4.51} & 4.44 & \textbf{4.54} & 0.26 & 0.13 \\
Suno-v4.5 & 7.62 & 7.84 & 6.27 & 8.24 & 4.60 & 4.48 & 4.61 & 4.48 & \textbf{4.48} & 4.51 & 0.25 & 0.14 \\
Mureka-V7.6 & 7.45 & 7.73 & 6.35 & 8.07 & 4.41 & 4.24 & 4.39 & 4.24 & 4.17 & 4.29 & 0.27 & 0.28 \\
Udio-v1.5 & 7.52 & 7.69 & 6.18 & 7.98 & 4.12 & 3.92 & 4.08 & 3.88 & 3.82 & 3.97 & 0.23 & 0.25 \\
MiniMax-2.0 & \textbf{7.73} & \textbf{7.98} & 6.45 & \textbf{8.35} & 4.60 & 4.49 & 4.59 & 4.48 & 4.39 & 4.51 & 0.26 & 0.13 \\ \midrule
LeVo & 7.55 & 7.79 & 5.81 & 8.22 & 3.62 & 3.43 & 3.49 & 3.42 & 3.29 & 3.45 & 0.13 & 0.22 \\
YuE & 6.26 & 7.21 & 4.58 & 7.18 & 2.80 & 2.58 & 2.66 & 2.54 & 2.61 & 2.64 & 0.14 & 0.65 \\
DiffRhythm 2 & 7.23 & 7.58 & 6.37 & 7.88 & 4.00 & 3.81 & 4.03 & 3.82 & 3.67 & 3.87 & \textbf{0.32} & 0.28 \\
ACE-Step & 7.49 & 7.61 & 5.86 & 7.90 & 3.45 & 3.17 & 3.31 & 3.27 & 3.06 & 3.25 & 0.17 & 0.21 \\
\midrule
\textbf{HeartMuLa (Ours)} 
& 7.55 & 7.82 & 5.89 & 8.14 
& 4.52 & 4.36 & 4.54 & 4.35 & 4.24 & 4.48 
& 0.26 & \textbf{0.09} \\ \bottomrule
\end{tabular}
}
\end{table*}

\begin{table*}[htbp]
\centering
\caption{Objective evaluation results on the HeartBeats Benchmark (Chinese).}
\label{tab:heartmula_main_result_zh}
\setlength{\tabcolsep}{3.5pt} 
\resizebox{\textwidth}{!}{ 
\begin{tabular}{lcccccccccccc}
\toprule
\multirow{2}{*}{\textbf{Model}} & \multicolumn{4}{c}{\textbf{AudioBox} $\uparrow$} & \multicolumn{6}{c}{\textbf{SongEval} $\uparrow$} & \multicolumn{1}{c}{\textbf{Align} $\uparrow$} & \multicolumn{1}{c}{\textbf{Intel} $\downarrow$} \\
\cmidrule(lr){2-5} \cmidrule(lr){6-11} \cmidrule(lr){12-12} \cmidrule(lr){13-13}
 & CE & CU & PC & PQ & Coh. & Mus. & Mem. & Cla. & Nat. & Avg. & Tag-Sim & PER \\ \midrule
Suno-v5 & 7.59 & 7.80 & \textbf{6.44} & 8.24 & \textbf{4.66} & \textbf{4.55} & \textbf{4.65} & \textbf{4.57} & \textbf{4.49} & \textbf{4.58} & \textbf{0.29} & 0.23 \\
Suno-v4.5 & 7.60 & 7.83 & 6.12 & 8.29 & 4.62 & 4.51 & 4.61 & 4.54 & 4.46 & 4.55 & 0.27 & 0.24 \\
Mureka-V7.6 & 7.39 & 7.70 & 6.39 & 8.16 & 4.42 & 4.27 & 4.36 & 4.28 & 4.20 & 4.30 & 0.28 & 0.29 \\
Udio-v1.5 & 7.35 & 7.57 & 6.16 & 8.02 & 4.14 & 3.95 & 4.05 & 3.94 & 3.87 & 3.99 & 0.26 & 0.50 \\
MiniMax-2.0 & \textbf{7.66} & \textbf{7.86} & 6.36 & 8.34 & 4.59 & 4.49 & 4.54 & 4.48 & 4.40 & 4.50 & 0.27 & 0.19 \\ \midrule
LeVo & 7.63 & 7.73 & 6.06 & \textbf{8.37} & 3.43 & 3.30 & 3.21 & 3.26 & 3.18 & 3.28 & 0.16 & 0.26 \\
YuE & 6.84 & 7.33 & 5.24 & 7.56 & 3.19 & 2.97 & 3.01 & 2.99 & 2.98 & 3.03 & 0.16 & 0.50 \\
DiffRhythm 2 & 7.23 & 7.59 & 6.25 & 8.07 & 3.92 & 3.72 & 3.85 & 3.76 & 3.60 & 3.77 & \textbf{0.29} & 0.32 \\
ACE-Step & 7.60 & 7.67 & 6.10 & 8.14 & 3.92 & 3.59 & 3.73 & 3.70 & 3.51 & 3.69 & 0.20 & 0.24 \\
\midrule
\textbf{HeartMuLa (Ours)} 
& 7.46 & 7.78 & 5.97 & 8.12 
& 4.57 & 4.45 & 4.58 & 4.45 & 4.36 & 4.48 
& 0.24 & \textbf{0.12} \\ 
\bottomrule
\end{tabular}
}
\end{table*}

\begin{table*}[htbp]
\centering
\caption{Objective evaluation results on the HeartBeats Benchmark (Japanese).}
\label{tab:heartmula_main_result_jp}
\setlength{\tabcolsep}{3.5pt} 
\resizebox{\textwidth}{!}{ 
\begin{tabular}{lcccccccccccc}
\toprule
\multirow{2}{*}{\textbf{Model}} & \multicolumn{4}{c}{\textbf{AudioBox} $\uparrow$} & \multicolumn{6}{c}{\textbf{SongEval} $\uparrow$} & \multicolumn{1}{c}{\textbf{Align} $\uparrow$} & \multicolumn{1}{c}{\textbf{Intel} $\downarrow$} \\
\cmidrule(lr){2-5} \cmidrule(lr){6-11} \cmidrule(lr){12-12} \cmidrule(lr){13-13}
 & CE & CU & PC & PQ & Coh. & Mus. & Mem. & Cla. & Nat. & Avg. & Tag-Sim & PER \\ \midrule
Suno-v5 & 7.47 & 7.71 & \textbf{6.57} & 8.21 & \textbf{4.63} & \textbf{4.54} & 4.60 & \textbf{4.51} & \textbf{4.44} & \textbf{4.54} & \textbf{0.32} & 0.35 \\
Suno-v4.5 & 7.44 & 7.73 & 6.44 & 8.22 & 4.58 & 4.48 & 4.56 & 4.48 & 4.42 & 4.51 & 0.27 & 0.29 \\
Mureka-V7.6 & 7.52 & 7.77 & 6.39 & \textbf{8.34} & 4.52 & 4.38 & 4.49 & 4.37 & 4.32 & 4.42 & 0.29 & 0.30 \\
Udio-v1.5 & 7.45 & 7.54 & 5.72 & 8.01 & 4.17 & 3.93 & 4.02 & 3.92 & 3.86 & 3.98 & 0.25 & 0.37 \\
MiniMax-2.0 & \textbf{7.65} & \textbf{7.87} & 6.38 & \textbf{8.34} & 4.49 & 4.36 & 4.47 & 4.34 & 4.28 & 4.39 & 0.27 & 0.24 \\ \midrule
YuE & 5.79 & 7.14 & 3.88 & 7.25 & 2.67 & 2.43 & 2.45 & 2.39 & 2.46 & 2.48 & 0.20 & 0.62 \\
ACE-Step & 7.62 & 7.69 & 6.14 & 8.23 & 3.71 & 3.38 & 3.49 & 3.42 & 3.27 & 3.45 & 0.23 & 0.33 \\
\midrule
\textbf{HeartMuLa (Ours)} 
& 7.42 & 7.80 & 6.32 & 8.19 
& 4.62 & 4.47 & \textbf{4.62} & 4.47 & 4.38 & 4.51 
& 0.24 & \textbf{0.20} \\ 
\bottomrule
\end{tabular}
}
\end{table*}

\begin{table*}[htbp]
\centering
\caption{Objective evaluation results on the HeartBeats Benchmark (Korean).}
\label{tab:heartmula_main_result_ko}
\setlength{\tabcolsep}{3.5pt} 
\resizebox{\textwidth}{!}{ 
\begin{tabular}{lcccccccccccc}
\toprule
\multirow{2}{*}{\textbf{Model}} & \multicolumn{4}{c}{\textbf{AudioBox} $\uparrow$} & \multicolumn{6}{c}{\textbf{SongEval} $\uparrow$} & \multicolumn{1}{c}{\textbf{Align} $\uparrow$} & \multicolumn{1}{c}{\textbf{Intel} $\downarrow$} \\
\cmidrule(lr){2-5} \cmidrule(lr){6-11} \cmidrule(lr){12-12} \cmidrule(lr){13-13}
 & CE & CU & PC & PQ & Coh. & Mus. & Mem. & Cla. & Nat. & Avg. & Tag-Sim & PER \\ \midrule
Suno-v5 & 7.44 & 7.75 & 6.40 & 8.15 & 4.61 & \textbf{4.52} & \textbf{4.61} & \textbf{4.52} & \textbf{4.47} & \textbf{4.54} & 0.30 & 0.38 \\
Suno-v4.5 & 7.42 & 7.77 & 6.40 & 8.20 & \textbf{4.61} & 4.47 & 4.55 & 4.50 & 4.41 & 4.51 & 0.28 & 0.40 \\
Mureka-V7.6 & 7.34 & 7.71 & 6.27 & 8.12 & 4.52 & 4.37 & 4.50 & 4.35 & 4.31 & 4.41 & 0.28 & 0.49 \\
Udio-v1.5 & 7.20 & 7.37 & 5.61 & 7.79 & 3.91 & 3.71 & 3.83 & 3.61 & 3.63 & 3.74 & 0.20 & 0.57 \\
MiniMax-2.0 & \textbf{7.62} & \textbf{7.92} & 6.22 & \textbf{8.38} & 4.57 & 4.43 & 4.51 & 4.40 & 4.33 & 4.45 & 0.26 & 0.40 \\ \midrule
YuE & 5.21 & 6.98 & 3.61 & 7.12 & 2.68 & 2.43 & 2.55 & 2.41 & 2.44 & 2.50 & 0.14 & 0.56 \\
ACE-Step & 7.53 & 7.67 & 6.12 & 8.17 & 3.53 & 3.25 & 3.33 & 3.29 & 3.15 & 3.31 & 0.20 & 0.38 \\
\midrule
\textbf{HeartMuLa (Ours)} 
& 7.51 & 7.79 & 6.09 & 8.16 
& 4.56 & 4.45 & 4.56 & 4.43 & 4.34 & 4.47 
& 0.21 & \textbf{0.16} \\ 
\bottomrule
\end{tabular}
}
\end{table*}

\begin{table*}[htbp]
\centering
\caption{Objective evaluation results on the HeartBeats Benchmark (Spanish).}
\label{tab:heartmula_main_result_es}
\setlength{\tabcolsep}{3.5pt} 
\resizebox{\textwidth}{!}{ 
\begin{tabular}{lcccccccccccc}
\toprule
\multirow{2}{*}{\textbf{Model}} & \multicolumn{4}{c}{\textbf{AudioBox} $\uparrow$} & \multicolumn{6}{c}{\textbf{SongEval} $\uparrow$} & \multicolumn{1}{c}{\textbf{Align} $\uparrow$} & \multicolumn{1}{c}{\textbf{Intel} $\downarrow$} \\
\cmidrule(lr){2-5} \cmidrule(lr){6-11} \cmidrule(lr){12-12} \cmidrule(lr){13-13}
 & CE & CU & PC & PQ & Coh. & Mus. & Mem. & Cla. & Nat. & Avg. & Tag-Sim & PER \\ \midrule
Suno-v5 & 7.48 & 7.71 & 6.47 & 8.20 & 4.55 & 4.42 & 4.53 & 4.43 & 4.37 & 4.46 & \textbf{0.25} & 0.15 \\
Suno-v4.5 & 7.45 & 7.75 & 6.27 & 8.24 & \textbf{4.60} & \textbf{4.47} & \textbf{4.59} & \textbf{4.49} & \textbf{4.44} & \textbf{4.52} & 0.19 & 0.31 \\
Mureka-V7.6 & 7.33 & 7.57 & 6.45 & 8.05 & 4.32 & 4.11 & 4.23 & 4.11 & 4.06 & 4.17 & 0.25 & 0.24 \\
Udio-v1.5 & 7.31 & 7.45 & 5.95 & 7.85 & 4.00 & 3.79 & 3.86 & 3.77 & 3.67 & 3.82 & 0.17 & 0.37 \\
MiniMax-2.0 & \textbf{7.71} & \textbf{7.91} & 6.50 & \textbf{8.35} & 4.49 & 4.35 & 4.47 & 4.32 & 4.25 & 4.38 & 0.22 & \textbf{0.12} \\ \midrule
ACE-Step & 7.43 & 7.44 & 6.18 & 7.99 & 3.24 & 2.92 & 2.96 & 2.94 & 2.87 & 2.99 & 0.10 & 0.21 \\
\midrule
\textbf{HeartMuLa (Ours)} 
& 7.39 & 7.72 &  6.12 & 8.16 
& 4.46 & 4.25 & 4.46 & 4.29 & 4.22 & 4.34 
& 0.22 & 0.13 \\ 
\bottomrule
\end{tabular}
}
\end{table*}

As shown in Tables \ref{tab:heartmula_main_result_en} to \ref{tab:heartmula_main_result_es}, HeartMuLa demonstrates stable and competitive performance across all five languages. The most significant advantage of our model is lyric clarity, where it achieves the lowest Phoneme Error Rate (PER) in every language tested. For instance, in English (0.09) and Chinese (0.12), HeartMuLa outperforms top closed-source models like Suno-v5 and MiniMax-Music-2.0, proving that it generates words much more clearly without blurring.

Regarding musical quality, HeartMuLa maintains high and consistent SongEval \cite{yao2025songevalbenchmarkdatasetsong} scores, with its structural and naturalness ratings matching professional standards. Unlike many open-source baselines that struggle with non-English tasks, HeartMuLa shows no performance drop across different regions. It maintains steady style adherence and audio quality in Chinese, Japanese, Korean, and Spanish alike. Overall, HeartMuLa provides a balanced combination of high-quality music production and industry-leading lyric accuracy for a global audience.

\subsubsection{Subjective Evaluation}
In this subjective evaluation experiment, 9 listeners participated in the test. Each rater evaluated 20 randomly selected samples from each model, including 10 English samples and 10 Chinese samples. To ensure fairness, a double-blind procedure with randomized trial order was adopted to minimize potential biases. 

To refine the raw annotations, we employed the CrowdMOS framework to remove outliers and exclude annotators who did not listen to the full audio samples. Following common industry practice, all audio samples were loudness normalized to -14 dB LUFS prior to evaluation. 

The subjective evaluation focused on seven perceptual dimensions:
\begin{itemize}
    \item \textbf{Musicality}: The overall aesthetic quality and emotional expressiveness of the song.
    \item \textbf{Harmony}: The coherence and compatibility between vocals and accompaniment.
    \item \textbf{Structure}: The logical progression and organization of the musical sections.
    \item \textbf{Fidelity}: The perceived audio quality and absence of artifacts.
    \item \textbf{Creativity}: The novelty and originality of the composition and performance.
    \item \textbf{Memorability}: The catchiness and long-term recall of the melody.
    \item \textbf{Text Alignment}: The semantic and rhythmic consistency between the generated lyrics and the musical content.
\end{itemize}

Participants were instructed to rate each stimulus according to the above criteria. The collected valid annotations were aggregated to compute the Mean Opinion Score (MOS) for each dimension.

All reported MOS values are computed as trimmed means by removing the highest and lowest ratings, with 95\% confidence intervals calculated based on the trimmed samples.

\begin{table*}[htbp]
\centering
\caption{Subjective Evaluation Results (Trimmed MOS $\pm$ 95\% CI) Across Different Dimensions}
\label{tab:mos_results}
\renewcommand{\arraystretch}{1.2}
\resizebox{\textwidth}{!}{%
\begin{tabular}{lccccccc}
\toprule
\textbf{Model} & \textbf{Musicality} & \textbf{Harmony} & \textbf{Structure} & \textbf{Fidelity} & \textbf{Creativity} & \textbf{Memorability} & \textbf{Text Align.} \\
\midrule
Suno-v4.5 &
78.10 $\pm$ 3.12 &
75.14 $\pm$ 4.16 &
78.80 $\pm$ 3.19 &
79.14 $\pm$ 2.65 &
71.26 $\pm$ 3.86 &
72.95 $\pm$ 3.81 &
77.17 $\pm$ 3.42 \\

\midrule
ACE-Step &
67.42 $\pm$ 4.40 &
67.16 $\pm$ 5.10 &
69.07 $\pm$ 4.31 &
71.40 $\pm$ 3.76 &
62.04 $\pm$ 4.91 &
61.60 $\pm$ 5.21 &
67.94 $\pm$ 4.44 \\

DiffRhythm 2 &
57.99 $\pm$ 5.38 &
57.13 $\pm$ 5.36 &
61.63 $\pm$ 4.62 &
62.58 $\pm$ 3.82 &
54.77 $\pm$ 5.06 &
53.15 $\pm$ 5.27 &
61.13 $\pm$ 4.45 \\

YuE &
56.95 $\pm$ 4.88 &
54.46 $\pm$ 5.37 &
60.07 $\pm$ 4.46 &
60.81 $\pm$ 3.99 &
55.68 $\pm$ 4.69 &
56.29 $\pm$ 4.80 &
61.25 $\pm$ 4.15 \\

HeartMuLa (Ours) &
69.55 $\pm$ 4.60 &
71.06 $\pm$ 4.57 &
73.44 $\pm$ 3.99 &
73.18 $\pm$ 3.74 &
66.73 $\pm$ 4.52 &
65.06 $\pm$ 4.93 &
70.51 $\pm$ 4.09 \\

\bottomrule
\end{tabular}
}
\end{table*}

\begin{table}[h]
\centering
\caption{Overall MOS Ranking Across All Subjective Metrics with 95\% Confidence Interval}
\label{tab:overall_mos}
\begin{tabular}{lccc}
\toprule
\textbf{Rank} & \textbf{Model} & \textbf{Overall MOS} & \textbf{95\% CI} \\
\midrule
1 & Suno-v4.5               & 76.08 & $\pm$1.33 \\
2 & HeartMuLa (Ours)   & 69.93 & $\pm$1.66 \\
3 & ACE-Step           & 66.66 & $\pm$1.76 \\
4 & DiffRhythm2         & 58.33 & $\pm$1.86 \\
5 & YuE                & 57.93 & $\pm$1.76 \\
\bottomrule
\end{tabular}
\end{table}

\subsubsection{Ablation Study}

\paragraph{The Effect of Different Training Phases.} To understand the contribution of each training phase, we conduct an ablation study on the three stages of HeartMuLa: Pretraining, Supervised Fine-Tuning (SFT), and Direct Preference Optimization (DPO). The results are summarized in Table \ref{tab:heartmula_ablation_study}.

\begin{table*}[htbp]
\centering
\caption{Ablation study (English) of different training stages.}
\label{tab:heartmula_ablation_study}
\setlength{\tabcolsep}{4pt} 
\resizebox{\textwidth}{!}{ 
\begin{tabular}{lcccccccccccc}
\toprule
\multirow{2}{*}{\textbf{Stage}} & \multicolumn{4}{c}{\textbf{AudioBox} $\uparrow$} & \multicolumn{6}{c}{\textbf{SongEval} $\uparrow$} & \multicolumn{1}{c}{\textbf{Align} $\uparrow$} & \multicolumn{1}{c}{\textbf{Intel} $\downarrow$} \\
\cmidrule(lr){2-5} \cmidrule(lr){6-11} \cmidrule(lr){12-12} \cmidrule(lr){13-13}
 & CE & CU & PC & PQ & Coh. & Mus. & Mem. & Cla. & Nat. & Avg. & Tag-Sim & PER \\ \midrule
HeartMuLa (Pretrain) & 7.15 & 7.61 & 5.68 & 7.86 & 4.16 & 3.95 & 4.12 & 4.12 & 4.04 & 4.03 & 0.2377 & 0.1830 \\
HeartMuLa (SFT) & 7.45 & 7.78 & 6.04 & 8.07 & 4.52 & 4.36 & 4.54 & 4.38 & 4.25 & 4.41 & 0.2499 & {0.1005} \\
HeartMuLa (DPO) & 7.58 & 7.81 & 5.94 & 8.10 & 4.53 & 4.40 & 4.56 & 4.37 & 4.27 & 4.43 & 0.2573 & 0.0687 \\ \bottomrule
\end{tabular}
}
\end{table*}

\begin{table*}[htbp]
\centering
\caption{Ablation study (Chinese) of different training stages.}
\label{tab:heartmula_ablation_study}
\setlength{\tabcolsep}{4pt} 
\resizebox{\textwidth}{!}{ 
\begin{tabular}{lcccccccccccc}
\toprule
\multirow{2}{*}{\textbf{Stage}} & \multicolumn{4}{c}{\textbf{AudioBox} $\uparrow$} & \multicolumn{6}{c}{\textbf{SongEval} $\uparrow$} & \multicolumn{1}{c}{\textbf{Align} $\uparrow$} & \multicolumn{1}{c}{\textbf{Intel} $\downarrow$} \\
\cmidrule(lr){2-5} \cmidrule(lr){6-11} \cmidrule(lr){12-12} \cmidrule(lr){13-13}
 & CE & CU & PC & PQ & Coh. & Mus. & Mem. & Cla. & Nat. & Avg. & Tag-Sim & PER \\ \midrule
HeartMuLa (Pretrain) 
& 7.19 & 7.62 & 5.77 & 7.89 
& 4.24 & 4.03 & 4.19 & 4.09 & 3.96 & 4.10 
& 0.2311 & 0.1174 \\
HeartMuLa (SFT) 
& 7.51 & 7.79 & 5.94 & 8.18
& 4.56 & 4.44 & 4.55 & 4.43 & 4.32 & 4.46
& 0.2442 & 0.1067 \\
HeartMuLa (DPO) 
& 7.49 & 7.78 & 5.97 & 8.18
& 4.52 & 4.37 & 4.50 & 4.38 & 4.26 & 4.40
& 0.2451 & 0.0806 \\
\bottomrule
\end{tabular}
}
\end{table*}

\paragraph{The Effect of DPO on Various Dimensions.} The experimental results shown in Table \ref{tab:DPO_results} demonstrate that Direct Preference Optimization (DPO) significantly enhances model performance across targeted dimensions, with each specialized preference dataset effectively optimizing its corresponding metric. Specifically, PER-DPO achieves a substantial reduction in Phoneme Error Rate (PER) to 0.0683, while Muq-DPO markedly improves semantic alignment, reaching a peak Tag-Sim score of 0.2839. Furthermore, the Audiobox \& Songeval-DPO model shows the most comprehensive gains in objective audio quality and subjective song evaluation, yielding the highest average score of 4.448. These findings validate the efficacy of dimension-specific DPO in refining music generation and the model we ultimately adopted was obtained by linearly merging the models trained on the three dimensions mentioned above.

\begin{table*}[htbp]
\centering
\caption{DPO results trained on three preference datasets including Muq-similarity-based, PER-based and Audiobox\&Songeval-based datasets.}
\label{tab:DPO_results}
\resizebox{\textwidth}{!}{
\renewcommand{\arraystretch}{1.2} 
\begin{tabular}{lccccccccccccc}
\toprule
\multirow{2}{*}{\textbf{Model}} & \multicolumn{4}{c}{\textbf{AudioBox} $\uparrow$} & \multicolumn{6}{c}{\textbf{SongEval} $\uparrow$} & \multicolumn{1}{c}{\textbf{Align} $\uparrow$} & \multicolumn{1}{c}{\textbf{Intel} $\downarrow$} \\ 
\cmidrule(lr){2-5} \cmidrule(lr){6-11} \cmidrule(lr){12-12} \cmidrule(lr){13-13}
 & CE & CU & PC & PQ & Coh. & Mus. & Mem. & Cla. & Nat. & Avg. & Tag-Sim & PER \\ 
\midrule
Base (SFT) & 7.45 & 7.78 & \textbf{6.04} & 8.07 & 4.52 & 4.36 & 4.54 & 4.38 & 4.25 & 4.41 & 0.2499 & 0.1005 \\
PER-DPO & 7.57 & 7.83 & 5.92 & 8.14 & 4.52 & 4.36 & 4.55 & 4.37 & 4.27 & 4.41 & 0.2538 & \textbf{0.0683} \\
Muq-DPO & 7.50 & 7.78 & 5.91 & 8.07 & 4.52 & 4.36 & 4.54 & 4.36 & 4.25 & 4.41 & \textbf{0.2839} & 0.0760 \\
Aud. \& Song.-DPO & \textbf{7.57} & \textbf{7.84} & 6.00 & \textbf{8.15} & \textbf{4.55} & \textbf{4.40} & \textbf{4.59} & \textbf{4.40} & \textbf{4.29} & \textbf{4.45} & 0.2636 & 0.0820 \\ 
\bottomrule
\end{tabular}%
}
\end{table*}


\subsection{Inference Acceleration}

Long-form music generation poses significant challenges for efficient inference, as it requires modeling long-range musical structure while preserving fine-grained acoustic details over thousands of autoregressive steps.
Under such conditions, inference latency is dominated mainly by repeated attention computation, Python-side scheduling overhead, and excessive GPU kernel launches.
To address these issues, we introduce a set of collaborative system-level optimizations for HeartMuLa, targeting the batch size $=1$ regime that is most relevant for interactive and streaming music generation. We employ a combination of KV-cache alignment, FlashAttention, and CUDA Graph to achieve efficient single-sample inference for long-form music generation, focusing on practical latency reduction and maintaining musical quality.

As summarized in Table~\ref{tab:heartmula_inference_performance}, the proposed optimizations jointly reduce end-to-end generation time from 398.3s in the baseline configuration to 73.4s, achieving a $5.4\times$ overall speedup.
At the same time, the total number of GPU kernel launches is reduced from 1,561,161 to 979,149, representing a substantial decrease that directly alleviates Python-side dispatch overhead and GPU launch latency, which are critical bottlenecks in autoregressive decoding.
The observed kernel launch reduction is slightly sub-linear relative to the latency reduction due to remaining Python-side scalar operations and dynamic control flow that limit the full reuse of kernels.

These gains are enabled by carefully co-designing the inference pipeline around HeartMuLa's \textbf{cascaded decoding architecture}, where a Global Transformer and a Local Transformer advance autoregressively in a tightly coupled manner~\cite{yang2023uniaudio}.
While this architecture improves modeling capacity and temporal coherence, it also introduces non-trivial challenges for inference efficiency, including strict state synchronization, frequent cache updates, and complex control flow across modules.

All optimizations described in this section are verified to preserve musical quality.
Objective evaluation results under different inference settings are summarized in Table~\ref{tab:heartmula_quality_full}, demonstrating that accelerated inference does not degrade generation quality.
Minor fluctuations across configurations are within expected variance and do not affect perceptual fidelity. Streaming inference achieves the lowest end-to-end latency and the best intelligibility, reflecting the effectiveness of KV-cache alignment and tensor-only control flow in maintaining temporal coherence.

\subsubsection{Overview of the Inference Pipeline}

HeartMuLa performs autoregressive generation over discrete acoustic tokens.
At each decoding step, the model predicts one frame consisting of $K$ RVQ code indices, conditioned on all previously generated frames and optional multimodal inputs such as lyrics or genre prompts.
Inference proceeds in two stages: \textbf{context prefill}, which initializes hidden states and KV-caches using the prompt sequence, and \textbf{autoregressive decoding}, which incrementally generates new acoustic frames.

During autoregressive decoding, the model produces one acoustic frame per step, making the pipeline naturally compatible with online and streaming generation.
Intermediate tokens or decoded audio segments can be emitted immediately once available.
In this work, we focus on batch size $=1$, which reflects practical interactive usage scenarios where end-to-end latency, responsiveness, and temporal stability are more important than throughput-oriented efficiency.

\subsubsection{Collaborative Optimizations: KV-Cache, FlashAttention, and CUDA Graph}

The primary computational bottleneck in long-form music generation arises from repeated self-attention over an ever-growing token history.
In cascaded decoding, this challenge is further exacerbated by the presence of multiple decoding modules whose internal states must remain strictly synchronized.

In the baseline implementation, frequent Python-side scalar operations (e.g., \texttt{.item()}) and dynamic control flow break computation graphs and lead to misaligned KV-cache updates.
As a result, KV-cache reuse becomes a \textbf{state consistency problem} rather than a simple memory optimization, contributing directly to the high latency and excessive kernel launches observed in the baseline setting (Tables~\ref{tab:heartmula_inference_performance}).

\paragraph{KV-Cache Alignment.}
To resolve these issues, we enforce strict alignment between token indices, positional encodings, attention masks, and KV-cache write locations.
At decoding step $t$, the KV-cache for each layer contains key-value pairs corresponding to tokens ${0, \ldots, t-1}$, and the newly generated token is appended at position $t$:

\begin{equation}
\mathcal{K}^{(l)}_{t+1} = \text{Append}(\mathcal{K}^{(l)}_t, k^{(l)}_t), \quad
\mathcal{V}^{(l)}_{t+1} = \text{Append}(\mathcal{V}^{(l)}_t, v^{(l)}_t).
\end{equation}

All cache updates are implemented using pure tensor operations, completely avoiding Python-side indexing or scalar extraction.
Positional encodings are derived from a global step counter, and separate positional buckets are used across modules when execution order differs.
This design guarantees consistency among attention masks, cache contents, and positional information, enabling effective KV-cache reuse and the latency reduction observed when KV-cache is enabled.

\paragraph{FlashAttention.}
To further accelerate self-attention, we integrate FlashAttention~\cite{dao2022flashattention, dao2023flashattention2} with KV-cache reuse.
Only the valid prefix of cached keys and values is exposed to the attention kernel, and causal masks are precomputed outside the decoding loop.
FlashAttention significantly reduces the number of kernel launches while maintaining stable speedups as sequence length increases, making it particularly effective for long-form music generation.

\paragraph{CUDA Graph.}
CUDA Graph is employed to reduce Python-side overhead by capturing and replaying static GPU execution graphs.
We carefully separate static and dynamic components of the inference pipeline: the core transformer forward passes are captured within the graph, while dynamic elements such as sampled tokens, updated positions, and control decisions are injected externally before replay.
This separation preserves shape invariance and correctness while enabling the substantial latency reduction reported in Table~\ref{tab:heartmula_inference_performance}.

\paragraph{Engineering Considerations.}
Several practical constraints limit the applicability of whole-graph compilation in cascaded decoding:

\begin{enumerate}
\item Warm-up executions prior to graph capture may implicitly populate KV-caches, causing positional misalignment during replay.
\item Sampling operations (e.g., multinomial sampling) inside captured graphs freeze the random state, resulting in identical outputs across runs; therefore, sampling must remain outside the graph.
\item Sharing positional buckets between backbone and decoder modules was observed to introduce subtle phase and timbre drift, motivating the use of independent buckets.
\item Minor changes in prompt length or codebook configuration frequently invalidate previously captured graphs, necessitating re-capture.
\item Python-side scalar operations such as \texttt{.item()} consistently break graph capture and introduce cache inconsistencies.
\item Dynamic control flow arising from special decoding modes (e.g., temperature $=0$ or variable CFG scales) cannot be safely included in captured graphs.
\end{enumerate}

Collectively, these observations motivate a strict tensor-only control flow and a clear separation between static and dynamic elements within the inference pipeline.

\subsubsection{Batching and Streaming Considerations}

The proposed inference pipeline supports both offline batch generation and online streaming generation.
The optimizations described above are primarily designed for the batch size $=1$ setting, which dominates interactive and streaming use cases.
In this regime, end-to-end latency and temporal stability are more critical than raw throughput.

As batch size increases, optimizations such as KV-cache reuse and FlashAttention remain effective, but their relative benefits diminish due to increased inter-module synchronization overhead.

In contrast, the inference pipeline naturally supports streaming generation.
Autoregressive decoding proceeds frame by frame, allowing intermediate acoustic tokens or decoded audio chunks to be emitted immediately.
KV-cache alignment and tensor-only control flow ensure temporal consistency across streamed outputs, while CUDA Graph capture is applied only to static subgraphs that do not interfere with online token emission.
As a result, HeartMuLa achieves stable low-latency streaming inference without compromising output quality.

\subsubsection{Experimental Evaluation}

To verify that inference acceleration does not degrade musical quality, we evaluate HeartMuLa under different inference configurations on our Benchmark.
Table~\ref{tab:heartmula_quality_full} reports objective metrics including AudioBox, SongEval, alignment (Align), and intelligibility (PER).
Across base, optimized, streaming, and batch inference settings, the results remain highly consistent, indicating that the proposed optimizations preserve melody, harmony, rhythmic structure, and perceptual quality.
Streaming inference achieves the best intelligibility (PER=0.0778) and the lowest end-to-end latency (67.95,s), demonstrating that real-time generation is feasible without significant quality degradation.

\begin{table*}[htbp]
\centering
\caption{End-to-end inference latency and GPU kernel launches under different HeartMuLa inference configurations.}
\label{tab:heartmula_inference_performance}
\setlength{\tabcolsep}{6pt}
\resizebox{\textwidth}{!}{
\begin{tabular}{lcccc}
\toprule
\textbf{Inference Setting} 
& \textbf{Avg. Latency (s)} $\downarrow$
& \textbf{Min (s)} $\downarrow$
& \textbf{Max (s)} $\downarrow$
& \textbf{Kernel Launches} $\downarrow$ \\
\midrule
Baseline                & 398.27 & 133.99 & 748.88 & 1{,}561{,}161 \\
KV-cache                & 132.99 & 85.04  & 171.95 & 987{,}887 \\
CUDA Graph              & 78.32  & 48.03  & 105.74 & 1{,}043{,}455 \\
FlashAttention          & 76.00  & 48.16  & 99.56  & 979{,}149 \\
All optimizations       & 73.41  & 46.24  & 98.88  & 979{,}149 \\
Streaming               & \textbf{67.95} & \textbf{45.27} & \textbf{89.07} & 979{,}149 \\
Batch (BS=1)            & 73.41  & 46.24  & 98.88  & 979{,}149 \\
Batch (BS=2)            & 116.02 & 62.00  & 124.01 & 942{,}473 \\
Batch (BS=4)            & 210.00 & 82.50  & 210.07 & 940{,}980 \\
Batch (BS=8)            & 390.06 & 248.75  & 423.50 & 939{,}510 \\
\bottomrule
\end{tabular}
}
\end{table*}

\begin{table*}[htbp]
\centering
\caption{HeartMuLa generation quality and AudioBox/SongEval metrics under different inference configurations.}
\label{tab:heartmula_quality_full}
\resizebox{\textwidth}{!}{%
\renewcommand{\arraystretch}{1.2}
\begin{tabular}{lcccccccccccc}
\toprule
\multirow{2}{*}{\textbf{Inference Setting}}
& \multicolumn{4}{c}{\textbf{AudioBox} $\uparrow$}
& \multicolumn{6}{c}{\textbf{SongEval} $\uparrow$}
& \multicolumn{1}{c}{\textbf{Align} $\uparrow$}
& \multicolumn{1}{c}{\textbf{Intel} $\downarrow$} \\
\cmidrule(lr){2-5} \cmidrule(lr){6-11} \cmidrule(lr){12-12} \cmidrule(lr){13-13}
 & CE & CU & PC & PQ & Coh. & Mus. & Mem. & Cla. & Nat. & Avg. & Tag-Sim & PER \\
\midrule
Baseline          & 7.478 & 7.789 & 5.901 & 8.167 & 4.574 & 4.445 & 4.579 & 4.441 & 4.348 & 4.477 & 0.2337 & 0.1008 \\
KV-cache          & \textbf{7.620} & \textbf{7.843} & 6.056 & \textbf{8.269} & \textbf{4.625} & \textbf{4.507} & \textbf{4.645} & \textbf{4.512} & \textbf{4.420} & \textbf{4.542} & 0.2440 & 0.0868 \\
CUDA Graph        & 7.576 & 7.822 & 5.932 & 8.211 & 4.550 & 4.429 & 4.561 & 4.431 & 4.334 & 4.461 & 0.2258 & 0.1235 \\
FlashAttention    & 7.557 & 7.812 & 6.084 & 8.209 & 4.610 & 4.491 & 4.601 & 4.482 & 4.410 & 4.519 & 0.2422 & 0.1053 \\
All optimizations & 7.585 & 7.818 & 5.982 & 8.224 & 4.615 & 4.481 & 4.602 & 4.490 & 4.400 & 4.518 & \textbf{0.2520} & 0.1351 \\
Streaming         & 7.606 & 7.830 & \textbf{6.111} & 8.255 & 4.618 & 4.486 & 4.609 & 4.481 & 4.392 & 4.517 & 0.2273 & \textbf{0.0778} \\
\bottomrule
\end{tabular}%
}
\end{table*}

\section{HeartCLAP}
The proposed HeartCLAP model architecture is designed to bridge the gap between musical audio and diverse textual descriptions. It comprises two primary components: a text encoder $E_T$ and a music encoder $E_M$. Both backbones are initialized with pre-trained weights from MuQ-MuLan \cite{zhu2025muq} to leverage robust prior musical knowledge. The high-level features extracted by both encoders are projected into a 1024-dimensional shared embedding space through separate linear projection layers.

To achieve cross-modal alignment, we employ Contrastive Learning using the InfoNCE loss \cite{oord2018representation}:

\begin{equation}
\begin{aligned}
    \mathcal{L}_\text{CLAP}=&-\sum^N_{i=1}\log\frac{\exp(s(m_i,t_i)/\tau)}{\exp(s(m_i,t_i)/\tau)+\sum^N_{j=1,j\neq i}\exp(s(m_i,t_j)/\tau)}\\
&-\sum^N_{i=1}\log\frac{\exp(s(m_i,t_i)/\tau)}{\exp(s(m_i,t_i)/\tau)+\sum^N_{j=1,j\neq i}\exp(s(m_j,t_i)/\tau)},
\end{aligned}
\end{equation}

where $s(\cdot, \cdot)$ denotes cosine similarity, $\tau$ is a learnable temperature parameter, and $m_i, t_i$ denotes the embedding extracted from the $i$-th music-text pair via $E_M$ and $E_T$ respectively. This objective ensures that positive music-text pairs $(m_i, t_i)$ are pulled together while mismatched pairs are pushed apart in the latent space.

\subsection{Training Details}
To foster a comprehensive understanding of music, our training data encompasses a wide spectrum of attributes: genre, mood, instrumentation, singer timbre, gender, scene, and topic. We utilize a multi-format annotation strategy where each audio segment is paired with both tag-based descriptions separated by commas like \texttt{"mood: soft, warm, genre: pop, ..."}, \texttt{"soft, warm, pop, ..."}, and natural language sentences like \texttt{"The music features a male gender, with hiphop genre, the instrument is drum, suitable for dance and workout scene"}. During training, one description format is randomly sampled per iteration to improve the model's linguistic flexibility. To enhance robustness against incomplete user prompts-a common scenario in music generation-we implement two masking strategies:

\begin{itemize}
\item \textbf{Attribute-level masking}: Entire attribute categories (e.g., all mood tags) are dropped with a probability $p_a$.
\item \textbf{Tag-level masking}: Within a category containing multiple tags (e.g., several instruments), individual tags are randomly omitted with a probability $p_t$.
\end{itemize}

The model was trained on 1 million music clips and corresponding description pairs using 8 NVIDIA A100 GPUs. We use the AdamW optimizer with a learning rate of $1 \times 10^{-4}$ and a weight decay of $0.1$. We employed a global batch size of 1024 and an epoch of 50.

\subsection{Experiment Results}
We evaluated our HeartCLAP along with the baselines Laion-CLAP \cite{elizalde2023clap} and MuQ-MuLan \cite{zhu2025muq} on the WikiMT-X \cite{wu2025clamp3} benchmark using music-description pairs for the Music-Text Retrieval task, as shown in Table $\ref{tab:retrieval_comparison}$. We employed Recall@K (R@K) and mAP@10 as evaluation metrics. R@$K$ is $1$ if the target-value item occurs in the first $K$ retrieved items, and $0$ otherwise. Meanwhile, mAP@10 calculates the average precision across all queries within the top 10 retrieved results.

\begin{table}[htbp]
\centering
\caption{Comparative Evaluation of CLAP Models on WiKiMT-X Benchmark.}
\label{tab:retrieval_comparison}

\begin{tabular}{lcccccc}
\toprule
\multirow{2}{*}{\textbf{Model}} & \multicolumn{3}{c}{Text-to-Music} & \multicolumn{3}{c}{Music-to-Text} \\ 
\cmidrule(lr){2-4} \cmidrule(lr){5-7}
 & R@1 $\uparrow$ & R@10 $\uparrow$ & mAP@10 $\uparrow$ & R@1 $\uparrow$ & R@10 $\uparrow$ & mAP@10 $\uparrow$ \\ \midrule
Laion-CLAP & 0.71 & 6.92 & 2.17 & 1.01 & 5.39 & 2.08 \\
MuQ-MuLan & 2.24 & 12.11 & 4.70 & 1.62 & 9.47 & 3.36 \\
HeartCLAP & \textbf{4.37} & \textbf{16.80} & \textbf{7.59} & \textbf{2.85} & \textbf{14.35} & \textbf{5.51} \\ \bottomrule
\end{tabular}

\end{table}

The experimental results demonstrate that HeartCLAP significantly outperforms both Laion-CLAP and the backbone MuQ-MuLan across all evaluation metrics. This substantial margin in R@$K$ and mAP@10 suggests that our framework effectively facilitates a more precise alignment between music and multi-faceted textual descriptions. By leveraging pre-trained musical priors and incorporating diverse annotation formats-ranging from structured tags to natural language, HeartCLAP exhibits superior generalization capabilities in capturing both high-level semantic themes and fine-grained acoustic attributes. The consistent performance gain in both Text-to-Music and Music-to-Text retrieval tasks indicates that the learned representations are mutually discriminative and robust for bi-directional retrieval scenarios.

\section{HeartTranscriptor}
\label{heart_transcriptor}

Mainstream Automatic Speech Recognition (ASR) models, such as Whisper \cite{whisper}, perform well on standard speech tasks. However, these systems struggle with music-specific ASR. Interference from complex instrumental accompaniments often degrades performance. Furthermore, the phonetic nuances of singing vocals differ significantly from spoken language. To address this, we introduce HeartTranscriptor, a framework designed for robust lyric recognition. Our approach involves fine-tuning Whisper on a high-quality dataset of musical audio. We adapt the model by aligning predicted probability distributions with ground truth labels. At each time step $t$, the decoder generates a probability distribution across the vocabulary. We then optimize model parameters using a cross-entropy loss function. The loss function $\mathcal{L}_\text{CE}$ is defined as:

\begin{equation}
\mathcal{L}_\text{CE} = - \sum_{t=1}^{T} \log P(y_t | y_{<t}, \mathbf{x}; \theta)
\label{eq:loss_function}
\end{equation}

where $\mathbf{x}$ represents the input audio features,
$y_t$ is the target lyric token at time step $t$, $y_{<t}$ denotes the preceding ground truth tokens, and $\theta$ refers to the learnable model parameters. During training, we employ teacher forcing by conditioning on ground truth history. This objective maximizes the likelihood of generating the correct lyric sequence given the audio input.

\subsection{Dataset Construction Pipeline} \label{sec:data_pipeline}
To construct the dataset, we first selected multilingual songs, including Chinese, English, Korean, Japanese, Spanish, and various other languages. We then employed the Demucs model \cite{10096956} to decouple the mixed audio tracks into isolated vocal and accompaniment tracks to mitigate interference. Subsequently, the Whisper-Medium model was utilized to perform lyric recognition on isolated vocal tracks. We then calculate the word error rates (WER) or character error rates (CER) over the transcribed results from Whisper-Medium.
Crucially, to ensure data quality, we implemented a hierarchical filtering strategy: retaining Chinese and English samples with an error rate below 0.7, and other languages below 0.8. In summary, this pipeline provides a high signal-to-noise ratio training corpus. Fine-tuning our model on this curated data markedly improved overall recognition performance.

\subsection{Training Details}

Based on the pipeline described above, we constructed a large-scale refined dataset of approximately 100,000 verified songs. These were further segmented into standardized 30-second audio slices, resulting in a cumulative duration of approximately 7,000~hours. This segment length aligns with the input window constraints of the Whisper model. By adhering to this duration, the model fully leverages available contextual features during training.

For the model training phase, we selected Whisper-Medium as the base model and employed a full fine-tuning strategy to effectively capture the complex acoustic textures present in music audio. The computational environment consisted of 8 NVIDIA A100 (80G) GPUs. We set the learning rate to $1 \times 10^{-5}$, coupled with a $1000$-step warmup strategy to prevent gradient instability during early training. To mitigate overfitting, the weight decay was set to $0.01$, and the gradient clipping threshold was capped at $1.0$. The training batch size was set to $16$, with gradient accumulation steps utilized to increase the effective batch size to $16 \times 4 \times 8$. Through this refined experimental configuration, we successfully developed a high-performance Music Automatic Speech Recognition model.

\subsection{Experiment}

To evaluate the performance of our fine-tuned model, we selected two distinct benchmarks. The first is SSLD-200 from SongPrep~\cite{tan2025songpreppreprocessingframeworkendtoend}, which consists of 100 full-length songs in both Chinese and English. The second is our internal HeartBeats-ASR-Bench, curated to test fine-grained capabilities with 200 audio slices (under 30 seconds) for each of the five target languages: English, Chinese, Korean, Japanese, and Spanish. Additionally, we utilized the Demucs model to process both datasets, isolating clean vocals from the background accompaniment. To quantitatively assess the accuracy of recognition, we adopted Word Error Rate (WER) for English and Spanish, and Character Error Rate (CER) for Chinese, Japanese, and Korean.

\begin{table}[htbp]
  \centering
  \caption{Comparison of WER and CER across different models and datasets. }
  \label{tab:wer_comparison_full}
  \resizebox{\textwidth}{!}{
    \begin{tabular}{lrrrrrrr}
    \toprule
    & \multicolumn{2}{c}{SSLD-200} & \multicolumn{5}{c}{HeartBeats-ASR-Bench} \\
    \cmidrule(lr){2-3} \cmidrule(lr){4-8}
    Model & 100en & 100zh & 200en & 200zh & 200ko & 200ja & 200es \\
    \midrule
    Whisper-Small & 0.6802 & 0.5605 & 0.4172 & 0.3473 & 0.4612 & 0.5556 & 0.4391 \\
    Whisper-Medium & 0.4695 & 0.4744 & 0.2206 & 0.3045 & 0.2789 & 0.3405 & 0.3520 \\
    Whisper-Turbo & 0.4737 & 0.3726 & 0.3397 & 0.3020 & 0.2672 & 0.2630 & 0.3328 \\
    Whisper-Large-V3 & 0.3981 & 0.3724 & 0.2139 & 0.3316 & 0.1646 & 0.2146 & 0.2798 \\
    SongPrep & 0.3460 & 0.1884 & 0.2075 & 0.1670 & \multicolumn{1}{c}{-} & \multicolumn{1}{c}{-} & \multicolumn{1}{c}{-} \\
    FireRedASR-aed & \multicolumn{1}{c}{-} & \multicolumn{1}{c}{-} & 0.5176 & 0.3471 & \multicolumn{1}{c}{-} & \multicolumn{1}{c}{-} & \multicolumn{1}{c}{-} \\
    FireRedASR-llm & \multicolumn{1}{c}{-} & \multicolumn{1}{c}{-} & 0.3266 & 0.1679 & \multicolumn{1}{c}{-} & \multicolumn{1}{c}{-} & \multicolumn{1}{c}{-} \\
    Qwen3-Omni-30B-A3B-Captioner & \multicolumn{1}{c}{-}  & \multicolumn{1}{c}{-}  & 0.2610 & 0.2049 &  0.2394  &  0.2804 & 0.2320 \\
    \midrule
    \textbf{HeartTranscriptor} & \textbf{0.2816} & \textbf{0.1438} & \textbf{0.1873} & \textbf{0.1077} & \textbf{0.1042} & \textbf{0.1801} & \textbf{0.2151} \\
    \bottomrule
    \end{tabular}%
  }
\end{table}

As shown in Table~\ref{tab:wer_comparison_full}, HeartTranscriptor consistently achieves the lowest error rates across all datasets and languages, significantly outperforming both the baselines and other domain-specific models. It demonstrates exceptional robustness, particularly in processing full-length and multilingual songs.  In comparison, the baseline Whisper models generally exhibit mediocre performance, with only Whisper-Large-V3 showing relatively better results among the vanilla versions. Regarding other specialized models, while SongPrep achieves competitive error rates, its applicability is strictly limited to Chinese and English. Furthermore, the FireRedASR series demonstrates average performance and is subject to strict input duration constraints (60s for aed, 30s for llm). Similarly, even after filtering out duration-induced hallucinations (error rate $> 3$) from Qwen3-Omni-30B-A3B-Captioner, it still underperforms compared to our proposed method.





\section{Overall Training Datasets}
\label{sec:overalldata}

\subsection{Composition of the Dataset}
Our dataset consists of three parts: music with lyrics, instrumental music, and TTS datasets (LJSpeech~\cite{itoLJSpeechDataset2017}, LibriTTS~\cite{zen2019librittscorpusderivedlibrispeech} and GigaSpeech~\cite{chen2021gigaspeech}). To account for musicality, we employed Audiobox-Aesthetic~\cite{tjandra2025metaaudioboxaestheticsunified} and SongEval~\cite{yao2025songevalbenchmarkdatasetsong} to filter the dataset. To ensure the alignment between lyrics and music, we utilized HeartTranscriptor for Automatic Speech Recognition (ASR) to eliminate mismatched music-lyrics pairs. Additionally, we trained a baseline model using the entire dataset (e.g., a music model capable of generating coherent sound) and filtered the training data based on its Perplexity (PPL). Ultimately, we retained approximately 100,000 hours of high-quality training data.

\subsection{Music Style}
\label{music_style}
For the music style in HeartMuLa, user inputs can vary widely, such as tags (e.g., K-pop), descriptions (e.g., "A lively music"), and more. Imagine you are a user: when selecting a style, the simplest and most direct approach would be to click a tag, forming your input. Furthermore, tags and descriptions can be interchanged via a Large Language Model (LLM). Thus, we have chosen a tag-based style approach. By analyzing user demand across various networks, we classified the tags into categories such as gender, genre, instrument, mood, scene, singer timbre, topic, and region. Based on these categories, we designed prompts that enable our fine-tuned Qwen2.5-Omni~\cite{xu2025qwen25omnitechnicalreport} to understand the music style. The prompts are as follows:

\begin{tcolorbox}[colback=gray!5, colframe=gray!60!black, title=\textbf{Prompt Detail}, arc=2mm]
\begin{lstlisting}[basicstyle=\ttfamily\scriptsize, breaklines=true, columns=fullflexible, frame=none]
You are a music researcher, music analysis assistant. Below I will give you a music, Please analyze this music. gender of the singer, genre of the music (ej, rock, hiphop), instrument used, mood of the music, scene(Applicable Scenarios for Music, excluding KTV,concert, party), singer's timbre, topic(Themes of Music), region, each result is retained within 1-3 words. must be in English. (genre, mood, scene, singer timbre, topic, region) these keys need contain **multiple** values.). You return the following format and explain why these labels were chosen.
{    
    "gender": "",
    "genre": [],
    "instrument": [],
    "mood": [],
    "scene": [],
    "singer_timbre": [],
    "topic": [],
    "region": []
}

**Reasoning:**
\end{lstlisting}
\end{tcolorbox}

\subsection{Music Structure}
\label{music_structure}
Through extensive experiments, we found that performing structural analysis of music during the pre-training phase and incorporating it into the lyrics significantly enhances the structural coherence of the music during inference. We employed SongFormer~\cite{hao2025songformerscalingmusicstructure} to annotate the structure of the music. An example of a structurally annotated music piece is shown in the table below:

\begin{tcolorbox}[colback=gray!5, colframe=gray!60!black, title=\textbf{Music Structure}, arc=2mm]
\begin{lstlisting}[basicstyle=\ttfamily\scriptsize, breaklines=true, columns=fullflexible, frame=none]
[Chorus]
My chest is vibrating like a V8 motor
Spinning faster, getting hotter
Every beat is a piston stroke
Leaving the past up in smoke

[Verse]
Cold air intake, breathing deep
Awake right now, no time for sleep
The valves are open, the rhythm is true
Driving this chassis straight to you

[Prechorus]
Throttle wide open, floor to the mat
No looking back, no time for that
Ignition sparks the electric blue

[Chorus]
My chest is vibrating like a V8 motor
Spinning faster, getting hotter
Every beat is a piston stroke
Leaving the past up in smoke

[Bridge]
Overheating in the red zone
But I can't stop this skin and bone
From racing down this road alone

[Outro]
Cut the engine.
Fade to black.
\end{lstlisting}
\end{tcolorbox}

\subsection{Data-Finegrained Style Annotation}

To enable precise user control over intra-song generation, we propose a Fine-grained Style Annotation Pipeline. Building upon the structured lyrics generated in \ref{music_structure} as input, this pipeline employs our fine-tuned Qwen-2.5-omni multimodal large language model as its core analytical engine. During the annotation process, adhering to the principle of Audio Grounding-where raw audio signals serve as the ultimate ground truth-the model integrates textual structural information to perform multi-dimensional feature deconstruction on each Structure Unit. Specifically, the analysis encompasses three orthogonal dimensions: Dynamics \& Energy, Vocal \& Technique, and Style \& Vibe. The detailed fine-grained style annotation is illustrated in Box~\ref{box:Finegrained Style Annotation}.

\begin{tcolorbox}[colback=gray!5, colframe=gray!60!black, title=\textbf{Music Finegrained Style Annotation},label=box:Finegrained Style Annotation,arc=2mm]
\begin{lstlisting}[basicstyle=\ttfamily\scriptsize, breaklines=true, columns=fullflexible, frame=none]
[Intro]
[Subtle electronic pulse, atmospheric build, anticipatory mood, understated energy]
Green leaves heavy on the bough
The secret starts right here and now

[Verse]
[Steady rhythmic foundation, introspective vocal delivery, narrative progression, moderate intensity]
No colors bursting in the spring air
No petals dancing in the wind without a care
People walk by and they don't see the show
But deep inside the skin,the sweetness starts to grow
A mystery wrapped up in a humble disguise
Hidden away from all the prying eyes

[Prechorus]
[Gradual crescendo, rising vocal urgency, building anticipation, subtle harmonic shift]
It happens in the dark,unseen
Beneath the canopy of green
Waiting for the perfect time to be

[Chorus]
[Explosive melodic hook, triumphant vocal expression, high energy sustain, core thematic statement]
Oh,the flower blooms inside the fruit
A hidden treasure,quiet and mute
We don't need the applause or the light
To make something beautiful and bright
Sweetness saved for those who know the truth
The essence of an everlasting youth

[Bridge]
[Melodic contrast, reflective vocal tone, shift in perspective, introspective pause]
Let the others chase the butterfly
Let them fade beneath the summer sky
We hold our magic close to the core

[Outro]
[Gradual fade, lingering melodic motif, sense of resolution, gentle conclusion]
Open it up and see
The secret of the fig tree


\end{lstlisting}
\end{tcolorbox}

\subsection{HeartBeats-Benchmark}
\label{heartbeats_benchmark}

We developed HeartBeats-Benchmark using a "Human-in-the-Loop" curation strategy to rigorously evaluate music understanding. Drawing inspiration from the dataset construction protocols like YuE \cite{yue} and DiffRhythm \cite{diffrhythm}, we invited professional musicologists to design a comprehensive evaluation taxonomy. This benchmark aims to map acoustic signals to high-level semantics through a rigorously verified standard.

\textbf{Evaluation Dimensions.}
To ensure a holistic assessment, the experts structured the analysis into three macro-perspectives covering six granular dimensions:

\begin{itemize}
    \item \textbf{Acoustic Structure}: This category covers Musical Style, delineating genre classification and stylistic roots. It also assesses Instrumentation, identifying rhythmic foundations and harmonic components within the arrangement.
    \item \textbf{Content Semantics}: This aspect examines Vocal Texture, characterizing the singer's timbre and acoustic properties. It further analyzes Narrative Theme, extracting subject matter and emotional focus from the lyric-melody interplay.
    \item \textbf{Contextual Atmosphere}: We incorporate Emotional Valence to verify alignment between audio features and human affective cognition. Additionally, Usage Scenario categorizes the functional settings suitable for the track.
    
\end{itemize}

To enhance input diversity, we implemented a Random Dimension Dropout strategy. Specifically, for each data sample, we randomly mask two out of the six granular dimensions defined in our taxonomy. The tags from the remaining four dimensions are then extracted to the final input prompt sequence. 

To visualize this transformation, a representative sample of this expert-validated schema is visualized in Box~\ref{box:tags}, while Box~\ref{box:final_tags} demonstrates the final inputs generated after applying the stochastic dropout strategy.

\begin{tcolorbox}[colback=gray!5, colframe=gray!60!black, title=\textbf{Box: Tag example}, label=box:tags, arc=2mm]
\begin{lstlisting}[basicstyle=\ttfamily\small, breaklines=true, columns=fullflexible, frame=none]
{
    "Acoustic Structure": ["pop","Piano", "Strings", "acoustic guitar", "synthesizer"],
    "Content Semantics": ["Sweet", "Love",]
    "Contextual Atmosphere": ["joyful","Dating"]
}
\end{lstlisting}
\end{tcolorbox}

\begin{tcolorbox}[colback=gray!5, colframe=gray!60!black, title=\textbf{Box: Final Input Sequence (Post-Dropout)}, label=box:final_tags, arc=2mm]

\begin{lstlisting}[basicstyle=\ttfamily\small, breaklines=true, columns=fullflexible, frame=none]
[Pop, Piano, Strings, Acoustic Guitar, Synthesizer, Joyful, Sweet]
\end{lstlisting}
\end{tcolorbox}

To ensure the professionalism and multilingual adaptability of the dataset, we established strict screening criteria for the lyrics section. The dataset includes 30 English, 20 Chinese, 10 Japanese, 10 Korean, and 10 Spanish songs. As demonstrated in Box~\ref{box:lyrics}, clear structural markers (e.g., [Intro], [Verse]) are embedded to segment the textual stream. Finally, the expert panel conducted a "Blind Validation" of all tag combinations, ensuring the dataset serves as an unbiased "Gold Standard" for model evaluation.

\begin{tcolorbox}[colback=gray!5, colframe=gray!60!black, title=\textbf{Box: Lyrics example}, label=box:lyrics, arc=2mm]
\begin{lstlisting}[basicstyle=\ttfamily\small, breaklines=true, columns=fullflexible, frame=none]
[Intro]

[Verse]
In the frozen North where the winter bites
I hold a crimson seed through the endless nights
It is hard as stone, preserved in the cold
A memory of you that never grows old

[Prechorus]
You are down South where the spring rain falls
Gathering the harvest by the garden walls
Two different worlds, but the color is the same

[Chorus]
Red beans scattered across the miles
One brings tears and the other brings smiles
Yours are blooming in the warm embrace
Mine are hidden in a secret place
North and South, we plant the yearning deep
Promises that we are sworn to keep
...
\end{lstlisting}
\end{tcolorbox}

\section{Related Works}
\subsection{Audio Tokenizer}
Discrete audio representations play a central role in audio language models, and existing approaches can be broadly categorized into semantic tokenizers and acoustic tokenizers, each exhibiting distinct trade-offs in representation capacity and modeling efficiency.

Semantic tokenizers typically rely on representations extracted from large-scale self-supervised learning (SSL) models, such as HuBERT \cite{hubert} and WavLM \cite{wavlm}. These models are trained to capture high-level linguistic and phonetic abstractions, making their hidden representations particularly amenable to discretization via clustering or vector quantization. Prior studies have shown that such semantic tokens exhibit strong compatibility with language modeling objectives, leading to improved modeling efficiency and generation stability in audio language models \cite{borsos2023audiolm, glm4-voice, cosyvoice, semanticodec}. However, this advantage comes at the cost of substantial loss of fine-grained acoustic information. As a result, semantic token based systems typically require additional generative decoders, e.g. diffusion models \cite{ddpm} or flow-matching frameworks \cite{lipman2022flow} to reconstruct waveforms. These cascaded architectures significantly increase inference latency and system complexity, limiting their practicality for large-scale or real-time generation.

In contrast, acoustic tokenizers are derived from neural audio codec models that are explicitly optimized for waveform reconstruction. Such codecs generally consist of an encoder, a discrete quantizer, and a lightweight decoder, enabling efficient end-to-end audio reconstruction with low inference overhead. Representative models, including EnCodec \cite{encodec}, HiFi-Codec \cite{yang2023hifi}, and DAC \cite{dac}, have demonstrated strong reconstruction fidelity across speech, music, and general audio domains. Owing to their rich acoustic expressiveness. 

Among existing low-bitrate semantic-aware codecs, MimiCodec \cite{moshi}, MuCodec \cite{mucodec_muencoder} are most closely related to our work. However, MimiCodec focuses on speech data, and MuCodec has limited reconstruction performance.  In contrast, our work targets a low-bitrate, semantic-rich, and high-fidelity music tokenizer. This design enables stronger alignment with language modeling objectives while retaining high-fidelity acoustic reconstruction. 

\subsection{Music generation}
\textbf{Language Model Paradigm} With the advent of Large Language Models (LLMs) and their demonstrated reasoning and scaling capabilities \cite{llama3}, several works have adopted LLMs for end-to-end music generation \cite{musicgen,musiclm}. A common framework involves encoding audio into discrete token sequences via Vector Quantized Variational Autoencoders (VQ-VAEs) \cite{vq-vae}  or Residual Vector Quantization (RVQ) \cite{soundstream_rvq}, which are subsequently modeled autoregressively.

\textbf{Diffusion Models for music generation}. Recently, diffusion models \cite{ddpm,lipman2022flow} have been introduced for their proficiency in modeling audio/music data \cite{make-an-audio,make-an-audio2}. 

\textbf{Song Generation} Song generation, which entails creating coherent vocals with instrumental accompaniment, presents unique challenges. Pioneering work Jukebox \cite{jukebox} uses a hierarchical VQ-VAE and transformer to model long sequences. Subsequent efforts like SongCreator \cite{songcreator} and MusiCot \cite{musicot} employ dual-sequence models or coarse-grained style conditioning to improve vocal-accompaniment relationships and overall structure, though they remain susceptible to interference between tracks and limited token vocabulary. Alternative approaches, such as MelodyLM \cite{melodylm}, adopt multi-stage pipelines to generate vocals and accompaniment separately. Recent models like YuE \cite{yue} and SongGen \cite{songgen} operate on interleaved vocal and accompaniment token sequences. On the diffusion front, DiffRhythm \cite{diffrhythm} generates full songs in a continuous space. While industrial systems (e.g., Suno, Udio) show impressive results, their technical details are not public. 

\section{Conclusion}
We presented HeartMuLa, an open-sourced family of music foundation models that unifies music-text alignment, music tokenization, lyrics recognition, and controllable music generation. By introducing a hierarchical audio language modeling framework built on ultra-low-frame-rate music tokens, our approach enables efficient and coherent long-form music generation with fine-grained controllability. Extensive experiments demonstrate consistent improvements over existing codecs and song generation baselines in both reconstruction quality, generation performance and modeling efficiency. We hope HeartMuLa serves as a strong foundation for future research in music understanding and generation, and facilitates broader applications in creative music production.

\section{Ethics and Responsibility}
HeartMuLa family is designed as an open-sourced initiative to advance music intelligence research. Our model operates on a transformative paradigm; it learns statistical acoustic representations to generate novel musical compositions rather than reproducing copyrighted source material. By utilizing an exceptionally diverse training dataset, explicitly enriched with culturally and linguistically diverse music content (see Section \ref{sec:overalldata}), our model can innovate and create within various musical styles effectively, thereby contributing to human musical artistry and cultural heritage. 

However, responsible innovation requires more than just advanced architecture. We echo the advocacy of \cite{ma2024foundationmodelsmusicsurvey} for labeling AI-generated/assisted content, thereby ensuring accountability for both creators and listeners. Notably, we implement a watermarking model to ensure audio security and facilitate content authentication. We believe this practice is essential for establishing a standardized responsibility protocol in the rapidly evolving field of AI music generation.
\newpage

\section{Contributions}
\noindent\textbf{Core Contributors}
\begin{itemize}
    \item Yuxin Xie, \textit{Peking University}
    \item Yuguo Yin, \textit{Peking University}
    \item Zheyu Wang, \textit{Scale global holding, Airo music technology}
    \item Xiaoyu Yi,  \textit{Peking University}
    \item Gongxi Zhu, \textit{Scale global holding, Airo music technology}
    \item Xiaolong Weng,\textit{Scale global holding, Airo music technology}
    \item Zihan Xiong, \textit{Scale global holding, Airo music technology}
    \item Yingzhe Ma, \textit{Scale global holding, Airo music technology}
    \item Dading Chong, \textit{Peking University}
    \item Dongchao Yang, \textit{The Chinese University of Hong Kong}
\end{itemize}
\vspace{1em}
\noindent\textbf{Contributors}
\begin{itemize}
    \item Jinliang Liu, \textit{Scale global holding, Airo music technology}
    \item Zihang Huang, \textit{Scale global holding, Airo music technology}
    \item Jinghan Ru, \textit{Peking University}
    \item Rongjie Huang, \textit{The Chinese University of Hong Kong}
    \item Haoran Wan, Peixu Wang, Kuoxi Yu. \textit{Scale global holding, Airo music technology} \\

\item Helin Wang, Liming Liang, Xianwei Zhuang. \textit{Peking University} \\

\item Yuanyuan Wang, Haohan Guo, Dingdong Wang \textit{The Chinese University of Hong Kong} \\

\item Junjie Cao, Zeqian Ju, Songxiang Liu, Yuewen Cao. \textit{Independent Researcher} 
\end{itemize}




\vspace{1em}
\noindent\textbf{Advisors}
\begin{itemize}
    \item Yuexian Zou, \textit{Peking University}
    \item Heming Weng, \textit{Scale global holding, Airo music technology}
\end{itemize}
\vspace{1em}
\noindent\textbf{Corresponding author}
\begin{itemize}
    \item Dongchao Yang$^{*}$,
    \textit{The Chinese University of Hong Kong}
\end{itemize}

\noindent $^{*}$Corresponding to:
\texttt{dcyang@se.cuhk.edu.hk}

\bibliographystyle{plainnat}
\bibliography{references}

\clearpage
\appendix



\end{document}